\documentclass[amsmath,amssymb,aps,eqsecnum]{revtex4}
\usepackage[dvips]{graphicx}
 \usepackage{bm,bbm}
\usepackage{epsfig}

\begin{document}
\title{An Introduction to Quantum Entanglement:\\
             A Geometric Approach}
\author{Karol {\.Z}yczkowski$^{1,2}$ and Ingemar Bengtsson$^3$}

\affiliation {$^1$Jagiellonian University, Cracow, Poland} 
\affiliation{$^2$Center for Theoretical Physics, 
  Polish Academy of Siences Warsaw, Poland}

\affiliation{$^3$Fysikum, Stockholm University, Sweden}

 \date{June 27, 2006}

\begin{abstract}
          We present a concise introduction to 
          quantum entanglement. Concentrating
          on bipartite systems we review the separability criteria
          and measures of entanglement. We focus our attention on geometry
          of the sets of separable and maximally entangled states.
          We treat in detail the two-qubit system 
          and emphasise in what respect this case is a special one.
\end{abstract}

\maketitle

\medskip
\begin{center}
{\small e-mail: karol@cft.edu.pl \quad \ ingemar@physto.se}
\end{center}


\hfill Entanglement is not {\sl one} but rather 

\hfill {\sl the} characteristic trait of quantum mechanics.\\

\hfill {\itshape ---Erwin Schr{\"o}dinger} 

\medskip

\section{Preface }

These notes are based on a chapter of our book 
{\sl Geometry of Quantum States. An Introduction to quantum entanglement}
\cite{BZ06}.
The book is written 
at the advanced undergraduate level
for a reader familiar with the principles of quantum mechanics.
It is targeted first of all for readers who 
do not read the mathematical literature everyday, but 
we hope that students of mathematics and of the information sciences will 
find it useful as well, since they also may wish to learn about 
quantum entanglement.

Individual chapters of the book are to a large extent
independent of each other. For instance, we are tempted 
to believe that the last chapter might become a source of information 
on recent developments on quantum entanglement 
also for experts working in the field. Therefore we 
have compiled these notes, which aim to present 
an introduction to the subject
as well as an up to date review 
on basic features of quantum entanglement.  
In particular we analyse pure and mixed states of a bipartite system, 
discuss geometry of quantum entanglement, 
review separability criteria and entanglement measures.

Since the theory of quantum entanglement in multipartite systems
is still at the forefront of a thriving research, we
deliberately decided not to attempt to cover 
these fascinating issues in our book.
As the book was completed in March 2005, we 
thought that the phenomenon of quantum entanglement in bi--partite quantum 
systems was already well understood.
However, the past year has brought several 
important results concerning the bi--partite case as well.
Although we had no chance to review these recent achievements 
in the book, we provide in Appendix A
some hints concerning the recent literature on the subject.

All references to equations or the numbers of section refers
to the draft of the book. 
To give a reader a better orientation on the topics
covered we provide its contents in Appendix B. 
Some practical exercises related to geometry
are provided in Appendix C.

\section{Introducing entanglement}
\label{sec:intent}

Working in a Hilbert space
that is a tensor product of the form ${\cal H} =
{\cal H}_A\otimes {\cal H}_B$, we were really interested in only one of
the factors; the other factor played the role of an ancilla describing
an environment outside our control. 
Now the perspective changes:
we are interested in a situation where there are two masters. The fate
of both subsystems are of equal importance, although they may be sitting
in two different laboratories. 

The operations performed independently
 in the two laboratories are described using
operators of the form ${\Phi}_A\otimes {\mathbbm 1}$ and ${\mathbbm 
1}\otimes {\Phi}_B$ respectively, but due perhaps to past history,
the global state of the system may not be a product state. In general,
it may be described by an arbitrary density operator ${\rho}$
acting on the composite Hilbert space ${\cal H}$.

\index{states!EPR}  
The peculiarities of this situation were highlighted in 1935 by
Einstein, Podolsky, and Rosen \cite{EPR35}.
Their basic observation was that
if the global state of the system is chosen suitably then it is possible to
change---and to some extent to choose---the state assignment in laboratory 
$A$ by performing operations
in laboratory $B$. The physicists in laboratory $A$ will be
unaware of this until they are told, but they can check in
retrospect that the experiments they performed were consistent with
the state assignment arrived at from afar---even though there was an 
element of choice in arriving at that state assignment. Einstein's considered 
opinion was that "on one supposition we should ... absolutely 
hold fast: the real factual situation of the system ${\cal S}_2$ is 
independent of what is done with the system ${\cal S}_1$, which is spatially
separated from the former" \cite{Ei49}. Then we seem to be forced to the 
conclusion that quantum mechanics is an incomplete theory in the sense that
its state assignment does not fully describe the factual situation in
laboratory $A$. 

In his reply to EPR, Schr{\"o}dinger argued that in quantum mechanics 
"the best possible knowledge of a {\it whole} does not include the 
best possible knowledge of all its {\it parts}, even though they may be 
entirely separated and therefore virtually capable of being 'best possibly 
known'\ ".
Schr\"odinger's (1935) ``general confession'' consisted  
of a series of three papers \cite{Sc34}, \cite{Sc35}, \cite{Sc36}. 
He introduced the word {\it Verschr{\"a}nkung} to
describe this phenomenon, personally translated it into English as
{\it entanglement}, and made some striking observations about it. 
The subject then lay dormant for many years.

To make the concept of entanglement concrete, we recall that the state of 
the subsystem in laboratory $A$ is given by the partially traced density 
matrix ${\rho}_A= \mbox{Tr}_B{\rho}$. This need not be a pure state, even 
if ${\rho}$ itself is pure. In the simplest possible case, 
namely when both ${\cal H}_A$ and ${\cal H}_B$ are two dimensional, we find 
an orthogonal basis of four states that exhibit this property in an extreme 
form. This is the {\it Bell basis}, 
\begin{equation}
 |{\psi}^{\pm}\rangle = 
\frac{1}{\sqrt{2}}
\bigl( |0\rangle |1\rangle \pm |1\rangle|0\rangle \bigr) \hspace{12mm} 
|{\phi}^{\pm}\rangle = 
 \frac{1}{\sqrt{2}}
\bigl( |0\rangle |0\rangle \pm |1\rangle|1\rangle \bigr) \ .
\label{singlet} \end{equation}

\noindent The Bell states all have the property that
${\rho}_A = \frac{1}{2}{\mathbbm 1}$, which means that we know nothing at
all about the state of the subsystems---even though we have maximal
knowledge of the whole. At the opposite extreme we have product states 
such as $|0\rangle |0\rangle $ and so on; if the global state of the 
system is in a product state then ${\rho}_A$ is a projector and the 
two subsystems are in pure states of their own. Such pure product 
states are called {\it separable},
while all other pure states are {\it entangled}.
\index{states!separable}
\index{states!entangled}

Now the point is that if a projective measurement is performed in laboratory
$B$, corresponding to an operator of the form  ${\mathbbm 1}\otimes 
{\Phi}_B$, then the global state will collapse to a product state. 
Indeed, depending on what measurement that $B$ choses to perform,
and depending on its outcome, the state in laboratory $A$ can become any
pure state in the support of ${\rho}_A$. 
(This conclusion was drawn by Schr\"odinger
from his mixture theorem. He found it ``repugnant''.) Of course, 
if the global state was one of the 
Bell states to begin with, then the experimenters in laboratory  $A$
still labour under the assumption that their state is
 ${\rho}_A = \frac{1}{2}{\mathbbm 1}$, 
and it is clear that any measurement results in $A$ will
be consistent with this state assignment. Nevertheless it would seem as
if the real factual situation in $A$ has been changed from afar.

In the early sixties John Bell \cite{Be64}
was able to show that if we hold
fast to the locality assumption then there cannot exist a completion of
quantum mechanics in the sense of EPR; it is the meaning of the
expression ``real factual situation" that is at stake in
entangled systems. The idea is that if the quantum mechanical
probabilities arise as marginals of a probability distribution over some
kind of a set of real factual situations, then the mere existence of the
latter gives rise to inequalities for the marginal distributions that,
as a matter of fact, are disobeyed by the probabilities
predicted by quantum mechanics.
But at this point opinions diverge; some
physicists, including notably David Bohm, have not felt obliged to hold 
absolutely fast to Einstein's notion of locality. See John Bell's (1987) 
\cite{Be87} book for a sympathetic review of Bohm's arguments. Followers 
of Everett (1957) \cite{Ev57} on the other hand argue that what happened was 
that the system in $A$ went from being entangled with the system in $B$ 
to being entangled with the measurement apparatus in $B$, with no change 
of the real factual situation in $A$.

Bell's work caused much excitement in philosophically oriented circles; 
it seemed to put severe limits on the world view offered by physics.
For a thorough discussion of 
the Bell inequalities consult Clauser and Shimony (1978) \cite{CS78}; 
experimental tests, notably by Aspect et al. (1982) \cite{ADR82}, show that 
violation of the Bell inequalities does indeed occur in the laboratory. 
(Although loopholes still exist; see Gill (2003) \cite{Gi03}.)

In the early nineties the emphasis began to shift. Entanglement came to 
be regarded as a {\sl resource} that allows us to do certain otherwise 
impossible things. An early and influential example is that of 
{\it quantum teleportation}. Let us dwell on this a little. The task 
is to send information that allows a distant receiver to reconstruct the 
state of a spin 1/2 particle---even if the state is unknown to the sender. 
But since only a single copy of the state is available the sender is 
unable to figure out what the state to be ``teleported'' actually is. So 
the task appears impossible.
(To send information that 
allows us to reconstruct a given state elsewhere 
is referred to as teleportation in the science fiction literature, 
where it is usually assumed to be trivial for the sender to 
verify what the state to be sent may be.) 
\index{teleportation}  

The idea of teleporting a state that is not known at all  
is due to Bennett et al. (1993)   
\cite{BB93}.
 A solution is to prepare a composite system in the Bell state 
$|{\phi}^+\rangle $, and to share the two entangled subsystems between sender 
and receiver. Suppose that state to be sent is ${\alpha}|0\rangle + 
{\beta}|1\rangle $. At the outset the latter is uncorrelated to the former, 
so the total (unnormalized) state is 
\begin{eqnarray} |{\Psi}\rangle = \bigl( {\alpha}|0\rangle + {\beta}|1\rangle
\bigr)
\bigl( |0\rangle |0\rangle + |1\rangle|1\rangle \bigr)
 = \hspace{20mm} \nonumber \\ 
\ \\ 
= {\alpha}
|0\rangle |0\rangle |0\rangle + {\alpha}|0\rangle |1\rangle |1\rangle + 
{\beta}|1\rangle |0\rangle |0\rangle + {\beta}|1\rangle |1\rangle |1\rangle 
\ . \nonumber \end{eqnarray}

\noindent The sender controls the first two factors of 
the total Hilbert space, and the receiver controls the third. By means of 
a simple manipulation we rewrite this as  
\begin{eqnarray} \sqrt{2}|{\Psi}\rangle = |{\psi}^+\rangle
\bigl({\alpha}|1\rangle + 
{\beta}|0\rangle \bigr) + |{\psi}^-\rangle \bigl({\alpha}|1\rangle - 
{\beta}|0\rangle \bigr) + \nonumber \\
\ \\
+ |{\phi}^+\rangle \bigl({\alpha}|0\rangle + {\beta}|1\rangle \bigr)
+ |{\phi}^-\rangle \bigl({\alpha}|0\rangle - {\beta}|1\rangle \bigr) \ .
\nonumber 
\end{eqnarray}

\noindent The sender now performs a projective measurement in the four
dimensional 
Hilbert space at his disposal, such that the state collapses to one of the 
four Bell states. If the collapse results in the state $|{\phi}^+\rangle $ 
the teleportation is complete. But the other cases are equally likely, so 
the sender must send two classical bits of information to the receiver, 
informing him of the outcome of her measurement. Depending on the result 
the receiver then performs a unitary transformation (such that $|0\rangle 
\leftrightarrow |1\rangle $, if the outcome was $|{\psi}^+\rangle $) and the 
teleportation of the still unknown qubit is complete.
This is not 
a {\sl Gedanken experiment} only; it was first done 
in Innsbruck \cite{Innsbruck} and in Rome \cite{Rome}.

In the example of teleportation the entangled auxiliary system was used to 
perform a task that is impossible without it. It will be noted also that 
the entanglement was used up, in the sense that once the transmission has 
been achieved no mutual entanglement between sender and receiver remains. 
In this sense then entanglement is a resource, just as the equally abstract 
concept of energy is a resource. Moreover it has emerged that 
there are many interesting tasks for which entanglement can be used, including 
{\it quantum cryptography} and {\it quantum computing}.

If entanglement is a resource we naturally want to know how much of it
we have. As we will see it is by no means easy to answer this question,
but it is easy to take a first step in the situation when the global
state is a pure one. It is clear that there is no entanglement in a
product state, when the subsystems are in pure states too and the von
Neumann entropy of the partially traced state vanishes. It is also clear 
that maximally entangled pure state will lead to a partially traced density 
matrix that is a maximally mixed state. For the case of two qubits the 
von Neumann entropy then assumes its maximum value
$\ln{2}$, and the amount of entanglement in such a state is known as an 
{\it e-bit}. 
\index{e-bit}  
States that are neither separable nor maximally entangled
require more thought. 
Let us write a pure state in its Schmidt form 
$|\Psi\rangle=\cos \chi \,|00\rangle + \sin \chi \, |11\rangle $.
Performing the partial trace one obtains 
\begin{equation} {\rho}_A = 
{\rm Tr}_B|\Psi\rangle \langle \Psi| =
\left[ \begin{array}{cc} \cos^2{\chi} & 0 \\ 
0 & \sin^2{\chi} \end{array} \right] \ . \end{equation}

\noindent The {\it Schmidt angle} ${\chi}\in [0,\pi/4]$ parametrizes 
the amount of ignorance about the state of the subsystem, that is to 
say the amount of entanglement. 
A nice thing about it is that its 
value cannot be changed by local unitary transformations of the form 
$U(2)\otimes U(2)$. For the general case, when the Hilbert space has 
dimension $N \times N$, we will have to think more, 
and for the case when the global state is itself a mixed one much more 
thought will be required. 

At this stage entanglement may appear to be such an abstract
notion that the need to quantify it does not seem to be urgent but
then, once upon a time, ``energy"
must have seemed a very abstract notion indeed, and now there are
thriving industries whose role is to deliver it in precisely quantified
amounts. Perhaps our governments will eventually have special 
{\sl Departments of Entanglement}
 to deal with these things. But that is in the far future; 
here we will concentrate on a geometrical description of entanglement 
and how it is to be quantified.  

\section{Two qubit pure states: entanglement illustrated}
\label{sec:22pure}

Our first serious move will be to take a look (literally) at entanglement 
in the two qubit case.
Such a geometric approach 
to the problem was initiated by Brody and Hughston (2001) \cite{BH01}, 
and developed in \cite{KZ01,MD01,BBZ02,Le04b}.
 Our Hilbert space has four complex dimensions, so 
the space of pure states is ${\mathbb C}{\bf P}^3$. We can make a picture 
of this space along the lines of section 4.6. So we draw the 
positive hyperoctant of a 3-sphere and imagine a 3-torus sitting over each 
point, using the coordinates
\begin{equation} (Z^0, Z^1, Z^2, Z^3) = (n_0,\,  n_1e^{i{\nu}_1},\, 
n_2e^{i{\nu}_2}, \,  n_3e^{i{\nu}_3}) \ . \end{equation}

\noindent The four non-negative real numbers $n_0$ etc. obey 
\begin{equation} n^2_0 + n_1^2 + n_2^2 + n_3^2 = 1 \ . \end{equation}

\noindent To draw a picture of this set we use a gnomonic projection 
of the 3-sphere centered at 
\begin{equation} (n_0, n_1, n_2, n_3) = \frac{1}{2}(1, 1, 1, 1) \ .
\end{equation}

\noindent The result is a nice picture of the hyperoctant, consisting of a
tetrahedron 
centered at the above point, with geodesics on the 3-sphere appearing as 
straight lines in the picture. The 3-torus sitting above each interior point 
can be pictured as a rhomboid that is squashed in a position dependent way. 

Mathematically, all points in ${\mathbb C}{\bf P}^3$ are equal. In physics, 
points represent states, and some states are more equal than others. 
In chapter 7
 this happened because we singled out a 
particular subgroup of the unitary group to generate coherent states. Now 
it is assumed that the underlying Hilbert space is presented 
as a product of two factors in a definite way, and this singles out the 
orbits of $U(N)\times U(N) \subset U(N^2)$ for special attention. More 
specifically there is a preferred way of using the entries $\Gamma_{ij}$ 
of an $N \times N$ matrix as homogeneous coordinates. Thus any 
(normalized) state vector can be written as 
\begin{equation}
 |{\Psi}\rangle \: =\:  \frac{1}{\sqrt{N}}\sum_{i=0}^n\sum_{j=0}^n
\, \Gamma_{ij}|i\rangle |j\rangle \ . 
\label{SchmidtC2}
\end{equation} 

\noindent For two qubit entanglement $N = n+1 =2$, and it is agreed that 
\begin{equation} (Z^0, Z^1, Z^2, Z^3)
\equiv (\Gamma_{00}, \Gamma_{01}, \Gamma_{10}, \Gamma_{11})
 \ . \end{equation}

\noindent Let us first take a look at the separable states. For such states 
\begin{equation}
 |{\Psi}\rangle = \sum_{i=0}^n\sum_{j=0}^n (a_i|i\rangle )
(b_j|j\rangle ) \hspace{5mm} \Leftrightarrow \hspace{5mm} \Gamma_{ij} = a_ib_j 
\ . 
\label{15.9}
 \end{equation}

\noindent In terms of coordinates a two qubit case state is separable if and 
only if 
\begin{equation} Z^0Z^3 - Z^1Z^2 = 0 \ . 
\label{15.10}
 \end{equation}

\noindent We recognize this quadric equation from 
section  4.3.
It defines the Segre embedding of ${\mathbb C}{\bf P}^1\otimes 
{\mathbb C}{\bf P}^1$ into ${\mathbb C}{\bf P}^3$. Thus the separable states 
form a four real dimensional submanifold of the six real dimensional space 
of all states---had we regarded ${\mathbb C}{\bf P}^1$ as 
a classical phase space, this submanifold would have been enough to describe 
all the states of the composite system. 

What we did not discuss in chapter 4 is the fact that the 
Segre embedding can be nicely described in the octant picture. Eq. 
(\ref{15.10}) splits into two real equations:
\begin{equation} n_0n_3 - n_1n_2 = 0 \end{equation}
\begin{equation} {\nu}_1 + {\nu}_2 - {\nu}_3 = 0 
\ . \end{equation}

\noindent Hence we can draw the space of separable states as a two dimensional 
surface in the octant, with a two dimensional surface in the torus that sits 
above each separable point in the octant. The surface in the octant has an 
interesting structure, related to Fig. 4.6. In eq. 
(\ref{15.9}) we can keep the state of one of the subsystems fixed; say 
that $b_0/b_1$ is some fixed complex number with modulus $k$. Then 
\begin{equation} \frac{Z^0}{Z^1} = \frac{b_0}{b_1} \hspace{7mm} 
\Rightarrow \hspace{7mm} n_0 = kn_1 \end{equation}
\begin{equation} \frac{Z^2}{Z^3} = \frac{b_0}{b_1} \hspace{7mm} 
\Rightarrow \hspace{7mm} n_2 = kn_3 \ . \end{equation}

\noindent As we vary the state of the other subsystem we sweep out a curve in 
the octant that is in fact a geodesic in the hyperoctant (the intersection
between 
the 3-sphere and two hyperplanes through the origin in the embedding space).
In the gnomonic coordinates that we are using this curve will appear as a
straight line, so what we see when we look at how the separable states sit 
in the hyperoctant is a surface that is ruled by two families of straight
lines. 

\begin{figure}
        \centerline{ \hbox{
                \epsfig{figure=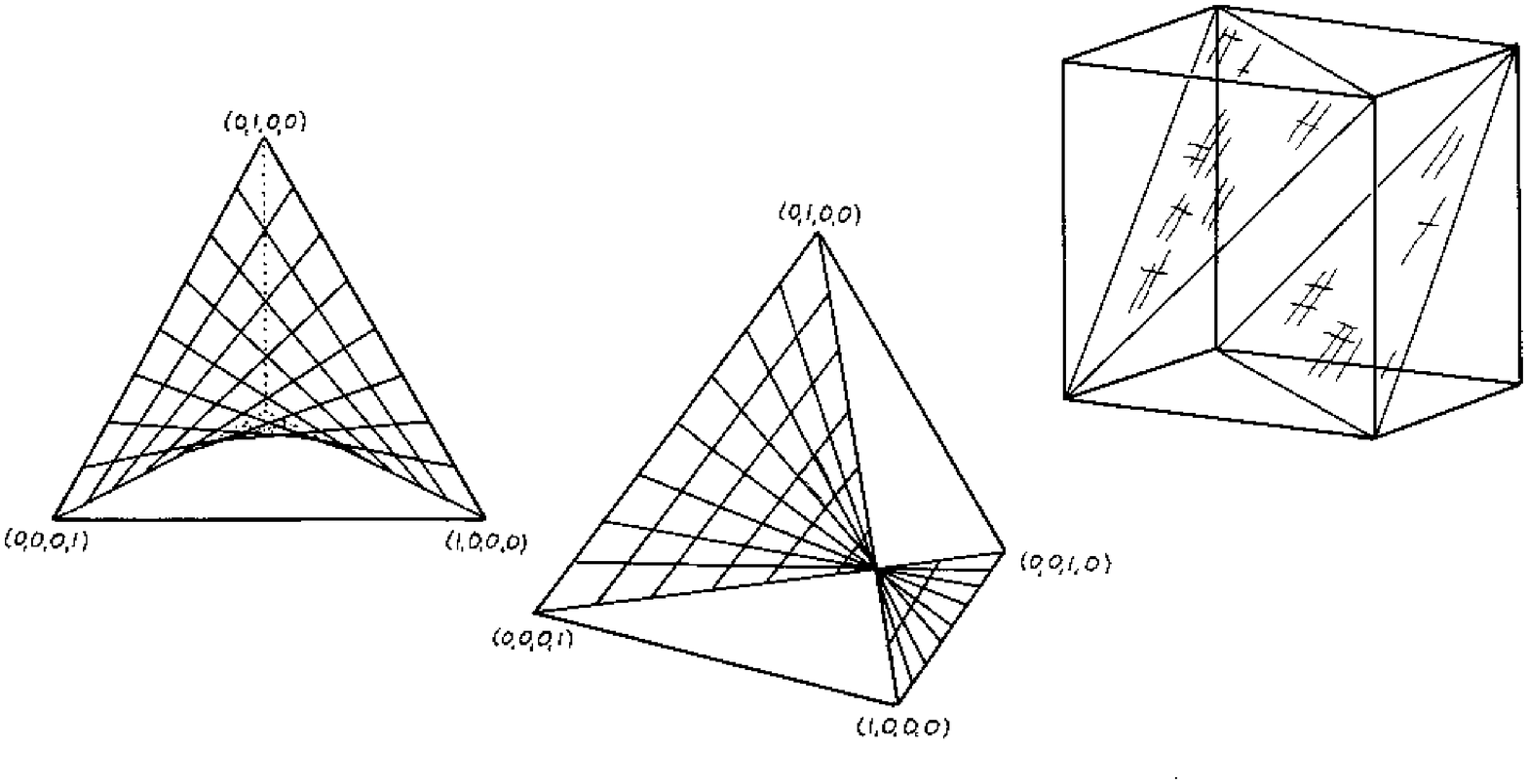,width=11.0cm}}}
        \caption{{\small The separable states, or the Segre embedding of 
${\mathbb C}{\bf P}^1\otimes {\mathbb C}{\bf P}^1$ in ${\mathbb C}{\bf P}^3$. 
Two different perspectives of the tetrahedron are given.}}
        \label{fig:Oktfig6}
\end{figure}

\index{Hopf fibration}  
There is an interesting relation to the Hopf fibration (see section 3.5)
 here. Each family of straight lines is also a one parameter 
family of Hopf circles, and there are two such families because there are 
two Hopf fibrations, with different twist. We can use our hyperoctant 
to represent real projective space ${\mathbb R}{\bf P}^3$, in 
analogy with Fig. 4.12. The Hopf circles that rule 
the separable surface are precisely those that get mapped onto each other 
when we ``fold'' the hemisphere into a hyperoctant. 
We now turn to the maximally entangled states, for which the reduced density 
matrix is the maximally mixed state. Using composite indices we write 
%
\begin{equation}
\rho_{\stackrel{\scriptstyle i j}{kl}}
 = \frac{1}{N}\Gamma_{ij}\Gamma^*_{kl} \hspace{5mm} 
\Rightarrow \hspace{5mm} {\rho}^A_{ik} = 
\sum_{j=0}^n 
\rho_{\stackrel{\scriptstyle i j}{kj}}\ . 
\end{equation} 
  Thus 
\begin{equation} {\rho}^A_{ik} = \frac{1}{N}{\mathbbm 1} \hspace{5mm} 
\Leftrightarrow \hspace{5mm} \sum_{j=0}^n \Gamma_{ij}\Gamma^*_{kj} =
{\delta}_{ik} \ . 
\end{equation}

\noindent Therefore the state is maximally entangled if and only if the 
matrix $\Gamma$ is unitary. Since an overall factor of this matrix is
irrelevant 
for the state we reach the conclusion that the space of maximally entangled 
states is $SU(N)/{\mathbb Z}_N$. This happens to be an interesting submanifold 
of ${\mathbb C}{\bf P}^{N^2-1}$, because it is at once {\it Lagrangian} 
(a submanifold with vanishing symplectic form and half the dimension of 
the symplectic embedding space) and {\it minimal} (any attempt to move it 
will increase its volume).

When $N = 2$ we are looking at $SU(2)/{\mathbb Z}_2 = 
{\mathbb R}{\bf P}^3$. To see what this space looks like in the octant 
picture we observe that 
\begin{equation}
 \Gamma_{ij} = \left[ \begin{array}{cc} {\alpha} & {\beta} \\ 
- {\beta}^* & {\alpha}^* \end{array} \right] \hspace{5mm} \Rightarrow 
\hspace{5mm} Z^{\alpha} = ({\alpha}, {\beta}, -{\beta}^*, {\alpha}^*) \ .
\label{CijZ}
 \end{equation}

\noindent In our coordinates this yields three real equations; the space 
of maximally entangled states will appear in the picture as a straight line 
connecting two entangled edges and passing through the center of the 
tetrahedron, while there is a two dimensional surface in the tori. The 
latter is shifted relative to the separable surface in such a way that 
the separable and maximally entangled states manage to keep their distance 
in the torus also when they meet in the octant (at the center 
of the tetrahedron where the torus is large). Our picture thus displays 
${\mathbb R}{\bf P}^3$ as a one parameter family of 
two dimensional flat tori, degenerating to circles at the ends of the 
interval. This is similar to our picture of the 3-sphere, except that this 
time the lengths of the two intersecting shortest circles on the tori 
stay constant while the angle between them is changing. It is amusing 
to convince oneself of the validity of this picture, and to verify that 
it is really a consequence of the way that the 3-tori are being squashed 
as we move around the octant. 

\begin{figure}
        \centerline{ \hbox{
                \epsfig{figure=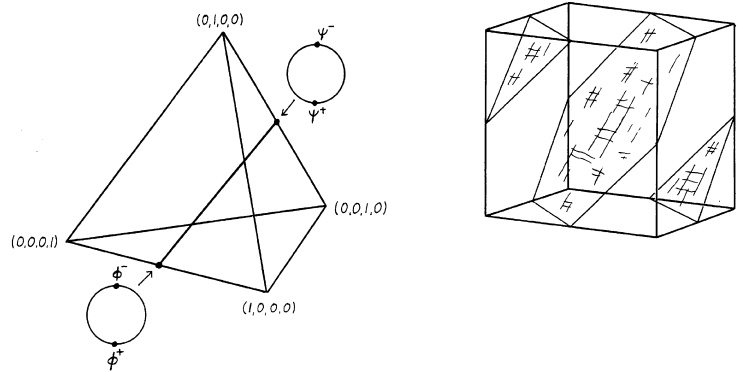,width=10cm}}}
        \caption{{\small The maximally entangled states form an ${\mathbb R}
{\bf P}^3$, appearing as a straight line in the octant and a surface in the 
tori. The location of the Bell states is also shown.}}
        \label{fig:Oktfig7}
\end{figure}

\begin{figure}[htbp]
        \centerline{ \hbox{
                \epsfig{figure=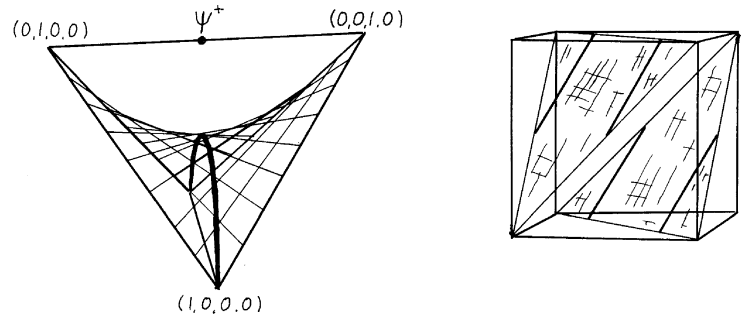,width=10cm}}}
        \caption{{\small A complete measurement on one of the subsystem
will collapse the Bell state $|{\psi}^+\rangle $ to a point on a sphere
on the separable surface; it appears as a one parameter family of circles 
in our picture. All points on this sphere are equally likely.}}
        \label{fig:Oktfig8}
\end{figure}

As a further illustration we can consider the collapse of a maximally 
entangled state, say $|{\psi}^+\rangle $ for definiteness, when a 
measurement is performed in laboratory $B$. The result will be a 
separable state, and because the global state is maximally entangled all 
the possible outcomes will be equally likely. It is easily confirmed that 
the possible outcomes form a 2-sphere's worth of points on the separable 
surface, distinguished by the fact that they are all lying on the same 
distance $D_{FS} = {\pi}/4$ from the original state. This is the minimal 
Fubini-Study distance between a separable and a maximally entangled state. 
The collapse is illustrated in Fig. \ref{fig:Oktfig8}.

A set of states of intermediate entanglement, quantified by some given value 
of the Schmidt angle ${\chi}$, is more difficult to draw (although it 
can be done). For the extreme cases of zero or one e-bit's worth of 
entanglement we found the submanifolds ${\mathbb C}{\bf P}^1\otimes 
{\mathbb C}{\bf P}^1$ and $SU(2)/{\mathbb Z}_2$, respectively. There is 
a simple reason why these spaces turn up, namely that the amount of 
entanglement must be left invariant under locally unitary transformations 
belonging to the group $SU(2)\times SU(2)$. In effect therefore we are 
looking for orbits of this group, and what we have found are the two 
obvious possibilities. More generally we will get a stratification of 
${\mathbb C}{\bf P}^3$ into orbits of $SU(2)\otimes SU(2)$; the problem 
is rather similar to that discussed in section 7.2. Of the 
exceptional orbits, one is a K\"ahler manifold and one (the maximally 
entangled one) is actually a Lagrangian submanifold of ${\mathbb C}{\bf P}^3$, 
meaning that the symplectic form vanishes on the latter. A generic orbit 
will be five real dimensional and the set of such orbits will be labeled 
by the Schmidt angle ${\chi}$, which is also the minimal distance from 
a given orbit to the set of separable states. A generic orbit is rather 
difficult to describe however. 
Topologically it is a non-trivial fibre bundle with an ${\bf S}^2$ as 
base space and ${\mathbb R}{\bf P}^3$ as fibre.
This can be seen 
in an elegant way using the Hopf fibration of ${\bf S}^7$---the space of 
normalized state vectors---as ${\bf S}^4 = {\bf S}^7/{\bf S}^3$; Mosseri 
and Dandoloff (2001) \cite{MD01} provide the details. In the octant
picture it appears as a three dimensional volume in the octant and a
two dimensional surface in the torus. 
And with this observation our tour of the two qubit Hilbert space 
is at an end. 

\section{Pure states of a bipartite system}
\index{entanglement}  
\index{pure!state}
\label{sec:NKpure}

Consider a pure state of a composite system 
$|\psi\rangle \in {\cal H}_{NK}={\cal H}_{N}\otimes {\cal H}_{K}$.
The states related by a local unitary transformation,
\begin{equation}
 |\psi' \rangle \ = \  U \otimes V \, |\psi\rangle \ ,
\label{locequiv}
 \end{equation}
\noindent 
where $U \in SU(N)$ and $V \in SU(K)$, 
are called {\it locally equivalent}. Sometimes one 
calls them {\it interconvertible states}, since they may be reversibly
converted by local transformations one into another \cite{JP99}.
It is clear that not all pure states are locally equivalent,
since the product group $SU(N)\times SU(K)$
forms only a measure zero subgroup of $SU(NK)$. 
How far can one go from a state using local transformations only?
In other words, what is  
the dimensionality and topology of the orbit
generated by local unitary transformations
from a given state $|\psi\rangle$?

To find an answer we are going to rely on the 
Schmidt decomposition (9.8). 
It consists of not more than $N$ terms, 
since without loss of generality we have assumed that $K\ge N$. 
The normalization condition $\langle \psi|\psi\rangle=1$ 
enforces $\sum_{i=1}^{N} \lambda_i=1$, 
so the Schmidt vector  ${\vec \lambda}$
lives in the ($N-1$) dimensional simplex $\Delta_{N-1}$.
\index{Schmidt!rank}  
The {\it Schmidt rank} of a pure state $|\psi\rangle$
is the number of  non--zero Schmidt coefficients,
equal to the rank of the reduced state.
States with maximal Schmidt rank are generic, and occupy the 
interior of the simplex, 
while states of a lower rank live on its boundary.

The Schmidt vector gives the spectra of the partially reduced states,
$\rho_A={\rm Tr}_B(|\psi\rangle \langle \psi |)$
and $\rho_B={\rm Tr}_A(|\psi\rangle \langle \psi |)$, 
which differ only by $K-N$ zero eigenvalues.
The separable states sit at the corners of the simplex. 
Maximally entangled states are described by the uniform Schmidt vector,
${\vec \lambda}_*=\{1/N,\dots,1/N\}$, since the partial trace
sends them into the maximally mixed state.

Let ${\vec \lambda}=(0,\cdots,0,\kappa_1,\cdots,\kappa_1,\kappa_2,
\cdots,\kappa_2,\dots,\kappa_J,\cdots,\kappa_J)$ 
represent an ordered Schmidt vector, 
in which each value $\kappa_n$ occurs $m_n$ times
 while $m_0$ is
the number of vanishing coefficients.
By definition $\sum_{n=0}^J m_n=N$, 
while $m_0$ might equal to zero.
The local orbit ${\cal O}_{\rm loc}$ generated 
from $|\psi\rangle$ has the structure
of a  fibre bundle, in which two quotient spaces

\begin{equation}
\frac {U(N)}{U(m_0)\times U(m_1)\times \cdots \times U(m_J)}
{\rm \quad and \quad} 
\frac {U(N)}{U(m_0)\times U(1)} 
 \label{locorbit}
\end{equation}
 form the base and the fibre, respectively \cite{SZK02}.
In general such a bundle need not be trivial.
The dimensionality of the local orbit may be computed
from dimensionalities of the coset spaces,
\begin{equation}
{\rm dim}({\cal O}_{\rm loc})= 2N^2-1-2m_0^2-\sum_{n=1}^J m_n^2 \ .
 \label{dimlocorb}
\end{equation}

\begin{table}[ht]
\caption{Topological structure of local orbits of the $N\times N$ \ pure
states,  
$D_s$ denotes the dimension of the subspace of the Schmidt simplex
$\Delta_{N-1}$,
while $D_o$ represents the dimension of the local orbit.}
\smallskip
{\renewcommand{\arraystretch}{1.45}
\begin{tabular}
[c]{cccccc}\hline 
$N$ & \parbox {2.5cm}{\centering Schmidt \\  coefficients} & $D_s$ 
 & \parbox {2.5cm}{\centering Part of the \\ asymmetric simplex} &
\parbox {2.5cm}{\centering Local structure:\\ base $\times$ fibre}
&  $D_o$
\\ \hline\hline
& $(a,b)$ & $1$ & line  & 
${\bf F}^{(2)}
\times
{\mathbb R}P^{3}$
 &
$5$
\\\cline{2-6}%
$2$ & $(1,0)$ & $0$ & left edge  & $
{\mathbb C}{\bf P}^{1}\times {\mathbb C}{\bf P}^{1}$ & $4$\\\cline{2-6}%
& $(1/2,1/2)$ & $0$ & right edge  & $U(2)/U(1)={\mathbb R}P^{3}$ &
$3$\\\hline\hline
& $(a,b,c)$ & $2$ & \parbox{3cm}{\centering 
interior of triangle} 
& $
{\bf F}^{(3)}
\times
 \frac{U(3)}{U(1)}$
 & $14$
\\\cline{2-6}%
& $(a,b,0)$ & $1$ & base & $
{\bf F}^{(3)}
\times
 \frac{U(3)}{[U(1)]^2}$ &
$13$
\\ \cline{2-6}
$3$ & $(a,b,b)$ & $1$ & 2 upper sides & 
$\frac{U(3)}{U(1) \times U(2)} 
\times
\frac{U(3)}{U(1)}$ 
 & $12$\\\cline{2-6}%
& $(1/2,1/2,0)$ & $0$ & right corner & 
$\frac{U(3)}{U(1) \times U(2)} 
\times
  \frac{U(3)}{[U(1)]^2}$ 
& $11$\\\cline{2-6}%
& $(1,0,0)$ & $0$ &  left corner  & $
{\mathbb C}{\bf P}^{2}\times {\mathbb C}{\bf P}^{2}$ & $8$%
\\\cline{2-6}%
& $(1/3,1/3,1/3)$ & $0$ &  upper corner & $U(3)/U(1)$ &
$8$\\\hline\hline
\end{tabular}
}
\label{tab:locorbit}
\end{table}
\begin{figure}[ht]
        \centerline{ \hbox{
                \epsfig{figure=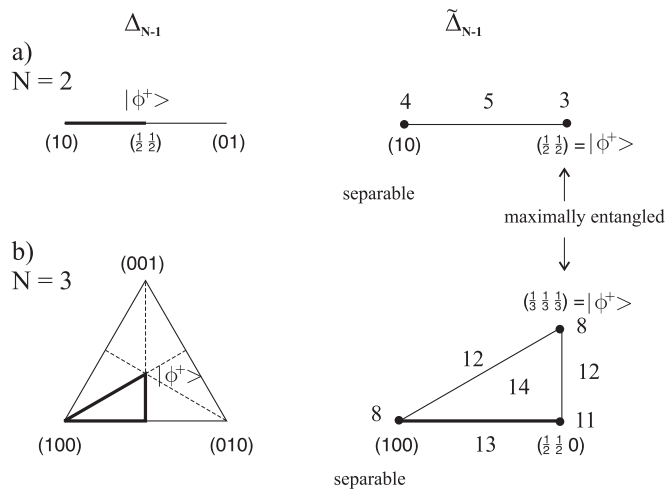,width=8.7cm}}}
        \caption{{\small Dimensionality of local orbits generated
              by a given point of the Weyl
    chamber $\tilde{\Delta}_{N-1}$ -- an asymmetric part
      of the Schmidt simplex $\Delta_{N-1}$ --   for pure
            states of $N \times N$ problem with $N=2,3$---compare 
        Tab. \ref{tab:locorbit}.}}
        \label{fig:ent01}
\end{figure}

Observe that the base describes the orbit of the
unitarily similar mixed states of the reduced system 
$\rho_A$, with spectrum $\vec \lambda$ and depends on its 
degeneracy -- compare Tab. 8.1. 
The fibre characterizes the manifold of pure states which
are projected by the partial trace to the same density matrix,
and depends on the Schmidt rank equal to $N-m_0$.
To understand this structure
consider first a generic state of the maximal Schmidt
rank, so that $m_0=0$. Acting on
$|\psi\rangle$ with $U_N \otimes W_N$, where both unitary matrices are
diagonal, 
we see that there exist $N$ redundant phases. Since each pure state is
determined up to an overall phase,  the generic orbit 
has the local structure  
\begin{equation}
{\cal O}_g \ \approx \ \frac {U(N)}{[U(1)]^N}
\ \times \
\frac {U(N)}{U(1)}=
 {\bf F}^{(N)} \ \times  \ \frac {U(N)}{U(1)},
\label{orbgen}
\end{equation}
with dimension ${\rm dim}({\cal O}_g) = 2N^2-N-1$.
If some of the coefficients are equal, say $m_J>1$, then we need to identify
all states differing by a block diagonal unitary rotation with $U(m_J)$ in the
right lower corner. In the same way one explains the meaning of the
factor $U(m_0)\times U(m_1)\times \cdots \times U(m_J)$ which appears in
the first quotient space of (\ref{locorbit}). 
If some Schmidt coefficients are equal to zero the action of the second
unitary matrix $U_B$ is trivial in the $m_0$---dimensional subspace---the 
second quotient space in (\ref{locorbit}) is $U(N)/[U(m_0)\times U(1)]$.

For separable states there exists only one non-zero coefficient, $\lambda_1=1$,
so $m_0=N-1$. 
\index{embedding!Segre} 
\index{manifold!flag} 
This gives the Segre embedding (4.16),  
\begin{equation}
 {\cal O}_{\rm sep}  = \frac {U(N)}{U(1)\times U(N-1)} \times
  \frac{U(N)}{U(1)\times U(N-1)}=
  {\mathbb C}{\bf P}^{N-1} \times {\mathbb C}{\bf P}^{N-1},
 \label{orbsep}
\end{equation}
of dimensionality ${\rm dim}({\cal O}_{\rm sep}) = 4(N-1)$. 
For a maximally entangled state
one has $\lambda_1=\lambda_N=1/N$, hence $m_1=N$
and $m_0=0$. Therefore
\begin{equation}
 {\cal O}_{\rm max}=\frac {U(N)}{U(N)} \times
  \frac{U(N)}{U(1)}= \frac{U(N)}{U(1)}=
  \frac{SU(N)}{{\mathbb Z}_N},
 \label{orbmax}
\end{equation}
with ${\rm dim}({\cal O}_{max}) = N^2-1$,
which equals {\sl half} the total dimensionality of
of the space of pure states.

The set of all orbits foliate ${\mathbb C}{\bf P}^{N^2-1}$, the space 
of all pure states of the $N \times N$ system. 
This foliation is {\it singular}, since 
there exist measure zero leaves of various 
dimensions and topology. The dimensionalities
of all local orbits for $N=2,3$ are shown 
in Fig. \ref{fig:ent01}, and their topologies 
in Tab. \ref{tab:locorbit}.

Observe that the local orbit defined by
 (\ref{locequiv}) contains all purifications
of all mixed states acting on ${\cal H}_N$
isospectral with 
$\rho_N={\rm Tr}_K|\psi\rangle \langle \psi|$.
Sometimes one modifies (\ref{locequiv}) imposing additional restrictions, $K=N$
and $V=U$. Two states fulfilling this 
{\it strong local equivalence} relation (SLE),
$|\psi' \rangle =  U \otimes U |\psi\rangle$
are {\sl equal}, up to selection of the reference frame used to 
describe both subsystems.
The basis are determined by a unitary $U$. 
Hence the orbit of the strongly locally equivalent states 
-- the base in  (\ref{locorbit})  -- forms a coset space 
of all states of the form $U\rho_N U^{\dagger}$.
 In particular, for any maximally entangled state,
there are no other states satisfying SLE,
 while for a separable state
the orbit of SLE states forms the complex projective space 
${\mathbb C}{\bf P}^{N-1}$ of all pure states of a single subsystem.

The question, if a given pure state $|\psi\rangle\in {\cal H}_N \otimes {\cal
H}_K$
is separable, is easy to answer: it is enough to compute the partial trace, 
$\rho_N={\rm Tr}_K(|\psi\rangle \langle \psi|)$, 
and to check if ${\rm Tr} \rho_N^2$  equals unity.
If it is so the reduced state is pure, hence the initial 
pure state is separable. In the opposite case the pure state is entangled.
The next question is: to what extent is a given state 
$|\psi\rangle$  entangled?

There seems not to be a unique answer to this question.
Due to the Schmidt decomposition
one obtains the Schmidt vector $\vec \lambda$
of length $N$ 
(we assume $N\le K$),
and may describe it by entropies analysed in chapters 2 and 12.
For instance, the {\it entanglement entropy} is defined as
the von Neumann entropy of the reduced state, 
which is equal to the Shannon entropy of the Schmidt vector,
\begin{equation}
E(|\psi\rangle) \  \equiv \ S(\rho_A)= S(\vec \lambda)
= -\sum_{i=1}^N \lambda_i \ln \lambda_i  \ .
\label{entanentr}
\end{equation}
\index{entropy!entanglement}
\noindent It is equal to zero for separable states and
$\ln N$ for maximally entangled states.
\index{entropy!R{\'e}nyi}  
In the similar way to measure entanglement one
may also use the R{\'e}nyi entropies (2.77) 
of the reduced state, $E_q\equiv S_q(\rho_A)$. 
We shall need a quantity related to $E_2$ called 
\index{concurrence}  
\index{tangle}  
{\it tangle}
\begin{equation}
\tau(|\psi\rangle) \ \equiv \ 
  2(1-{\rm Tr} \rho_A^2 )
= 2 \Bigl( 1-\sum_{i=1}^N \lambda_i^2 \Bigr)
= 2 \Bigl( 1-\exp[-E_2(|\psi\rangle)] \Bigr) \ ,
\label{tangle}
\end{equation}
which runs from $0$ to $2(N-1)/N$, and its square root $C=\sqrt{\tau}$, called 
{\it concurrence}.
Concurrence was initially introduced for two qubits
by Hill and Wootters  \cite{HW97}. 
We adopted here the generalisation of  \cite{RBCHM01,MKB04},
but there are also other ways to generalise
this notion for higher dimensions \cite{Uh00,WC01,Woo01,AVM01,BDHHH02}.

 Another entropy,
$E_{\infty}=-\ln \lambda_{\rm max}$, has a nice geometric  
interpretation: if the Schmidt vector is
ordered decreasingly and  $\lambda_{1}=\lambda_{\rm max}$
denotes its largest component
then $|1\rangle \otimes |1\rangle$
is the separable pure state closest to $|\psi\rangle$ \cite{LS02}.
Thus the Fubini--Study distance of $|\psi\rangle$ to the 
set of separable pure states,
$D_{\rm FS}^{\rm min}=\arccos(\sqrt{\lambda_{\rm max}})$, 
is a function of $E_{\infty}$.
\index{distance!Fubini-Study}  
Although one uses several different  R{\'e}nyi entropies $E_q$,
the entanglement entropy $E=E_1$ is distinguished among 
them just as the Shannon entropy is singled out by
its operational meaning 
discussed in section 2.2.

For the two qubit problem  the Schmidt vector
has only two components, which sum to unity, 
so the entropy $E(|\psi\rangle) \in [0,\ln 2]$
characterises uniquely the entanglement of the pure state $|\psi\rangle$.
To analyze its geometry it is convenient
to select a three dimensional section 
of the space of pure states - see Fig. \ref{fig:ent00}.
The net of this tetrahedron 
is shown in appendix 
it presents entanglement at the boundary
of the simplex defined by four separable
states  defining the standard basis.
\begin{figure}
        \centerline{ \hbox{
          \epsfig{figure=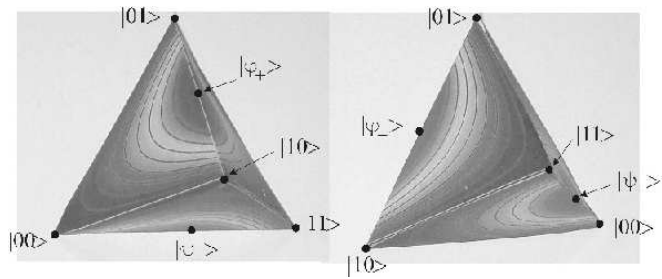,width=8.7cm}}}
        \caption{{\small Entropy of entanglement in gray scale
 for two $3$-D cross-sections of the $6$-D manifold of pure
   states of two qubits. Both tetrahedrons
show the same object from different angles --  compare also Figs.
\ref{fig:Oktfig6} and 
\ref{fig:Oktfig7}.}}
        \label{fig:ent00}
\end{figure}

In general, for an $N \times N$ system the entropy is bounded,
$0 \le E \le \ln N$, and to describe the entanglement completely one  needs a
set of $N-1$ independent quantities. What 
properties should they fulfill?

\index{operations!local} 
Before discussing this issue we need to distinguish certain classes of quantum
operations
acting on bipartite systems. 
{\it Local operations} (LO) arise as the tensor product of two maps,
both satisfying the trace preserving condition, 
\begin{equation}
[\Phi_A \otimes \Phi_B] (\rho)\: =\:  \sum_i \sum_j 
   (A_i \otimes B_j)\, \rho \, (A_i^{\dagger} \otimes B_j^{\dagger}) \ .
 \label{localoper}
\end{equation}
Any operation which might be written in the form
\begin{equation}
\Phi_{\rm sep}(\rho)\: =\: \sum_i  
   (A_i \otimes B_i)\, \rho \, (A_i^{\dagger} \otimes B_i^{\dagger}) \ ,
 \label{separoper}
\end{equation}
is called {\it separable} (SO).
\index{operations!separable} %
Observe that this form is more general than (\ref{localoper}),
even though the summation  goes over one index. 
\index{LOCC}    
\index{operations!LOCC} 
The third, important class of maps is called 
{\it LOCC}. This name stands for
{\it local operations and classical communication}
and means that all quantum operations, including
measurements, are allowed, provided they are performed
locally in each subsystem. Classical communication
allows the two parties to exchange 
in both ways classical information
about their subsystems, and hence to introduce
classical correlations between them. One could think, 
all separable operations may be obtained in this
way, but this is not true \cite{BVFMRSSW99},
and we have the proper inclusion relations 
${\rm LO}\subset {\rm LOCC} \subset {\rm SO}$.

The concept of local operations leads to the notion 
of {\it entanglement monotones}.
\index{entanglement!monotones} 
\index{local invariants} 
These are the quantities which
are invariant under unitary operations and decrease,
on average, under LOCC \cite{Vi00}. 
The words 'on average'
refer to the general case, in which a pure state is
transformed by a probabilistic local operation 
into a mixture, 
\begin{equation}
\rho \ \to \ \sum_i p_i \rho_i \ \ \Rightarrow \ \ 
  \mu (\rho) \ \ge \ \sum_i p_i \mu (\rho_i)  
   \ .
 \label{monot}
\end{equation}
Note that if $\mu$ is a non--decreasing monotone,
then $-\mu$ is a non--increasing monotone.
Thus we may restrict our attention
to the non--increasing monotones,
which reflect the key paradigm
of any  entanglement measure: entanglement cannot
increase under the action of local operations.
Construction of entanglement monotones
can be based on
\smallskip

\noindent
{\bf Nielsen's majorisation theorem} \cite{Ni99}.
{\sl A given state $|\psi\rangle$ may be transformed
into $|\phi\rangle$
\index{theorem!Nielsen} 
by  deterministic LOCC operations
 if and only if the
corresponding vectors of the Schmidt coefficients
satisfy the majorisation relation} (2.1)
\begin{equation}
|\psi\rangle \: 
\stackrel{\rm LOCC}{\longrightarrow} \: |\phi\rangle
\quad 
\Longleftrightarrow
\quad
{\vec \lambda}_{\psi} \: \prec \: {\vec \lambda}_{\phi} .
 \label{equipur}
\end{equation}

\noindent To prove the forward implication we follow the original proof.
Assume that party $A$ performs locally
a generalised measurement, which is
described by a set  of $k$ Kraus operators $A_i$.
By classical communication
the result is sent to party $B$, which
performs a local action $\Phi_i$,
conditioned on the result $i$. Hence
\begin{equation}
\sum_{i=1}^k [{\mathbbm 1} \otimes \Phi_i] \bigl( 
 A_i |\psi\rangle \langle \psi | A_i^{\dagger} \bigr) =
    |\phi\rangle \langle \phi| \ .
\label{niels1}
\end{equation}
The result is a pure state so each terms in the sum 
needs to be proportional to the projector.
Tracing out the second subsystem
we get 
\begin{equation}
A_i\rho_{\psi}A_i^{\dagger}=p_i \rho_{\phi} \ , \quad \quad i=1,\dots ,k,
\label{niels2}
\end{equation}
where $\sum_{i=1}^k p_i=1$ and
$\rho_{\psi}={\rm Tr}_B(|\psi\rangle \langle \psi|)$
and $\rho_{\phi}={\rm Tr}_B(|\phi\rangle \langle \phi|)$.
Due to the polar decomposition of $A_i \sqrt{\rho_{\psi}}$
we may write 
\begin{equation}
A_i \, \sqrt{\rho_{\psi}} =\sqrt{ A_i \rho_{\psi} A_i^{\dagger} } \ V_i =
       \sqrt{p_i \rho_{\phi}} \ V_i  
\label{niels3}
\end{equation}
with unitary $V_i$. Making use of the completeness relation 
we obtain  
\begin{equation}
\rho_{\psi}=\sqrt{\rho_{\psi}}\, {\mathbbm 1}\sqrt{\rho_{\psi}}
= \sum_{i=1}^k \sqrt{\rho_{\psi}} A_i^{\dagger} A_i \sqrt{\rho_{\psi}}
 =\sum_{i=1}^k p_i V_i^{\dagger} \rho_{\phi}  V_i,
\label{niels4}
\end{equation}
and the last equality follows from (\ref{niels3}) and its adjoint.
Hence we arrived at an unexpected conclusion:
if a local transformation $|\psi\rangle \to |\phi\rangle$
is possible, then there exists a bistochastic
operation (10.64), which acts on 
the partially traced states with inversed time -
it sends $\rho_{\phi}$ into $\rho_{\psi}$!
The quantum HLP lemma (section 2.2)
 implies the majorisation relation
${\vec \lambda}_{\psi} \prec {\vec \lambda}_{\phi}$.
The backward implication follows from
an explicit conversion protocol proposed by Nielsen,
or alternative versions presented in \cite{Har01,JSc01,DHR02}.

\index{majorisation} 
The majorisation relation (\ref{equipur})
introduces a partial
order into the set of pure states.
(A similar partial 
order induced by LOCC into the space of mixed states 
is analysed in \cite{HTU01}.)
 Hence any pure state $|\psi\rangle$
allows one to split the Schmidt simplex, representing the set of all local
orbits,
into three regions: the set $F$ ({\sl Future}) contains states
which can be produced from $|\psi\rangle$ by LOCC ,
the set $P$ ({\sl Past}) of states from which
$|\psi\rangle$ may be obtained, and eventually
the set $C$ of {\it incomparable} states, which 
cannot be joined by a local transformation in any direction.
For $N\ge 4$
there exists an effect of {\it  entanglement catalysis} \cite{JP99,DK01,BRo02}
that allows one to obtain certain incomparable states
in the presence of additional entangled states.

\begin{figure} [htbp]
   \begin{center}
\
 \includegraphics[width=11.0cm,angle=0]{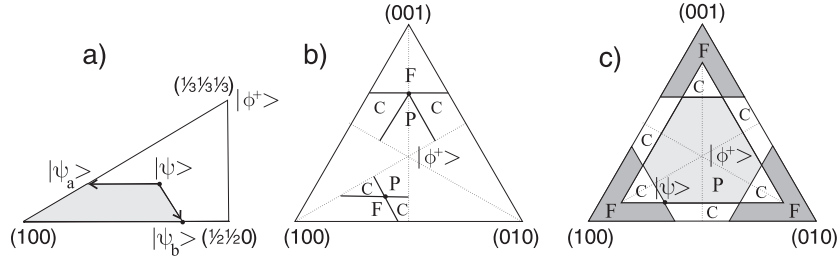}
\caption{
 Simplex of Schmidt coefficients $\Delta_2$ for $3\times 3$ pure states:
 the corners represent separable states,  center the maximally entangled state
 $|\phi^+\rangle$. Panels (a-c) show '{\sl Future}' and '{\sl Past}'
zones with respect to LOCC, 
} 
\label{fig:ent02}
\end{center}
 \end{figure}

This structure resembles the ``causal structure'' defined by the light 
cone in special  relativity. See  Fig. \ref{fig:ent02}, and observe the 
close similarity to figure 12.2
showing  paths in the simplex of eigenvalues that can 
be generated by bistochastic operations. 
The only difference is the {\it arrow of time}:
the 'Past' for the evolution in the space of density matrices
corresponds to the 'Future' for the local entanglement transformations
and vice versa. In both cases the set $C$ of incomparable states 
contains the same fragments of the simplex $\Delta_{N-1}$.
In a typical case $C$ occupies regions close to the boundary
of $\Delta_{N-1}$, so one may expect, the  larger dimensionality 
$N$,  the larger  relative volume of $C$. This is indeed the case,
and in the limit $N\to \infty$
two generic pure states of the $N\times N $ system
(or two generic density matrices of size $N$) are incomparable \cite{CHW02}. 

\index{Schmidt!simplex} 
The majorisation relation (\ref{equipur}) 
provides another justification for the observation 
that two pure states are interconvertible (locally equivalent) 
if and only if the have the same Schmidt vectors.
More importantly, this theorem implies  
that any Schur concave function 
of the Schmidt vector $\vec \lambda$ 
is an entanglement monotone.
In particular, this crucial  property 
is shared by all R{\'e}nyi entropies of entanglement $E_q(\vec \lambda)$
including the entanglement entropy (\ref{entanentr}). To 
ensure a complete description of a pure state of
the $N \times N$ problem one may choose
$E_1, E_2, \dots , E_{N-1}$.
\index{Schur convexity} 
Other families of entanglement monotones include 
partial sums of Schmidt coefficients ordered
decreasingly,  $M_k(\vec \lambda)=\sum_{i=1}^k \lambda_k$
with $k=1,\dots,N-1$ \cite{Vi00},
subentropy \cite{JRW94,MZ03},
and symmetric polynomials in Schmidt
coefficients.

Since the maximally entangled state 
is majorised by all pure states, it cannot be reached
from other states by any deterministic local transformation.
Is it at all possible to create it locally?
A possible clue is hidden in the word
{\sl average} contained in the majorisation theorem.

Let us assume we have at our disposal $n$ copies
of a generic pure state $|\psi\rangle$. The majorisation theorem
does not forbid us to locally create out of them $m$ maximally
entangled states $|\psi^+\rangle$, at the expense of
the remaining $n-m$ states becoming separable.
Such protocols
proposed in  \cite{BBPS96,LP01}
are called {\it entanglement concentration}. 
This local operation is reversible,
and the reverse process of transforming $m$ 
maximally entangled states and $n-m$
separable states into $n$ entangled states
is called {\it entanglement dilution}.
The asymptotic ratio $m/n \le 1$
obtained by an optimal concentration protocol
is called {\it distillable entanglement} \cite{BBPS96}
of the state $|\psi\rangle$.

Assume now that only one copy of an 
entangled state $|\psi\rangle$
is at our disposal. To generate 
maximally entangled state locally
we may proceed in a probabilistic way:
a local operation produces
$|\psi^+\rangle$ with probability $p$ and 
a separable state otherwise.
Hence we allow a pure state $|\psi\rangle$
to be transformed into a mixed state. 
Consider a probabilistic
scheme to convert a pure state
$|{\psi}\rangle$  into a target $|\phi\rangle$
with probability $p$.
Let $p_c$ be the maximal number such that
the following majorisation relation holds,
\begin{equation}
\lambda_{\psi} \ \prec \  
p_c  \, \vec \lambda_{\phi} .
 \label{submaj}
\end{equation}
It is easy to check that
the probability $p$ cannot be larger
than $p_c$, since the Nielsen theorem 
would be violated.
The optimal conversion strategy for which
$p=p_c$ was explicitly constructed by Vidal \cite{Vi99}.
The Schmidt rank cannot increase during
any local conversion scheme \cite{LP01}.  
If the rank of the target state $|\phi\rangle$
is larger than the Schmidt rank of $|{\psi}\rangle$,
then $p_c=0$ and the probabilistic conversion
cannot be performed.
In such a case one may
still perform a {\it faithful conversion} \cite{JP99a,VJN00}
 transforming the initial state 
 $|\psi \rangle$ into a state $|\phi ' \rangle$,
for which its fidelity with the target,  
$| \langle \phi|\phi' \rangle|^2$,  is maximal. 

\begin{figure} [ht] 
   \begin{center} \
 \includegraphics[width=11.0cm,angle=0]{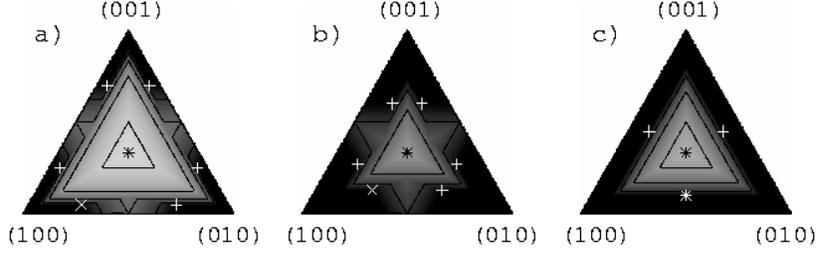}
\caption{Probability of the
optimal local conversion of an initial state $|\psi\rangle$\
(white $\times$) of the $3 \times 3$ problem 
 into a given pure state represented by a point in the Schmidt
simplex.
Initial Schmidt vector $\vec \lambda_{\psi}$ is 
 (a) $(0.7, 0.25, 0.05)$, (b) $( 0.6, 0.27,0.03)$,
 and (c) $(0.8, 0.1, 0.1)$.
Due to the degeneracy of $\vec \lambda$ in the latter case 
there exist only three interconvertible states in $\Delta_2$,
represented by 
$(+)$. 
} 
\label{fig:ent03}
\end{center}
 \end{figure}

This situation is illustrated  in Fig.  \ref{fig:ent03}, which 
shows the probability of accessing different regions of the
Schmidt simplex for pure states of a $3 \times 3 $ system for four
different initial states $|\psi\rangle$. The shape of the black figure
($p=1$ represents deterministic transformations) is identical with the set
'Future' in Fig. \ref{fig:ent02}. The more entangled final state $|\phi\rangle$
(closer to the maximally entangled state -- black $(*)$  in the center of the
triangle),
the smaller probability $p$ of a successful transformation.
Observe that the contour lines (plotted at $p=0.2,0.4,0.6$ and $0.8$)
are constructed from the iso-entropy lines $S_{q}$ for
$q\to 0$ and $q \to \infty$,

Let us close with an envoi:
entanglement of a pure state of any
bipartite system may be
fully characterized by its Schmidt
decomposition. In particular,
all entanglement monotones 
are functions of the Schmidt coefficients.
However, the Schmidt decomposition cannot
be directly applied to the multi--partite case \cite{Pe95,CHS00,AAJ01}. 
These systems are still being investigated -
several 
families of local invariants and entanglement monotones
were found  \cite{Su00,BC01,Gi02},
properties of local orbits were analysed
\cite{MD01,BC03,Mi03,Le04b}, 
measures of multi--partite entanglement
were introduced \cite{CKW00,BPRST01,WC01,MW02,Br03,HBj04},
and a link between quantum mechanical and topological entanglement
including  knots and braids \cite{KL02,AKMR04}, rings \cite{OW01},
and graphs \cite{PB03,HEB04} has been discussed.
Let us just mention that pure states of three 
qubits can be entangled in two inequivalent ways.
There exist three-qubit
 pure states \cite{GHZ89}
\begin{equation}
|GHZ\rangle =\frac{1}{\sqrt{2}} \bigl( |000\rangle +|111\rangle \bigr)
{\quad \rm and \quad}
|W\rangle =\frac{1}{\sqrt{3}} \bigl( |001\rangle +|010\rangle + |100\rangle
\bigr) \,
 \label{GHZ}
\end{equation}
which cannot be locally converted with a positive probability
in any direction \cite{DVC00}.
A curious
reader might be pleased to learn that {\sl four} qubits can be entangled in
{\sl nine} 
different ways \cite{VDDV02}.
What is the number of different ways, one may entangle $m$ qubits?

\section{Mixed states and separability}
\label{sec:sepmix}

It is a good time to look again at mixed states:
in this section we shall analyze bipartite density matrices,
acting on a composite Hilbert space ${\cal H}={\cal H}_A \otimes {\cal H}_B$
of finite dimensionality $d=NK$. 
A state is called a {\it product state}, if it has a tensor product structure,
$\rho=\rho_A\otimes \rho_B$.   
A mixed state $\rho$ is called {\it separable}, if it can be represented as
a convex sum of  product states \cite{We89}, 
\begin{equation}
\rho_{sep}\: = \: \sum_{j=1}^M \, q_j \, \rho^A_j \otimes \rho^B_j \ ,
 \label{mixedsep}
\end{equation}
where $\rho^A$ acts in ${\cal H}_A$ and $\rho^B$ acts in ${\cal H}_B$,
the weights are positive, $q_j>0$,  and  sum to unity, $\sum_{j=1}^M q_j=1$.
Such a decomposition is not unique.
 For any separable $\rho$,
the smallest number $M$ of terms 
is called {\it cardinality}\,
of the state.
Due to Carath\'eodory's theorem
the cardinality is not larger then $d^2$ \cite{Ho97}. In the two--qubit case 
it is not larger than $d=4$ \cite{STV98},
while for systems of higher dimensions 
it is typically larger than the rank $r \le d$ \cite{Lo00}.

By definition the set ${\cal M}_S$ of separable mixed states is convex.
Separable states can be constructed locally using classical communication,
and may exhibit classical correlations only.
A mixed state which is not separable, hence may display non--classical 
correlations,  is called {\it entangled}.
It is easy to see
that for pure states both definitions are consistent.
The notion of {\sl entanglement}
may also be used in the set-up of classical probability distributions
\cite{Tu00},
theory of Lie-algebras or convex sets \cite{BKOV03}
and may be compared with {\sl secret} classical correlations \cite{CP02}.

Any density matrix $\rho$ acting on $d$ 
dimensional Hilbert space may be represented 
as a sum (8.10) over $d^2-1$ trace-less generators 
$\sigma_i$ of $SU(d)$. 
However, analysing a composite system for which $d=NK$,
it is more advantageous to use the basis
of the product group $SU(N) \otimes SU(K)$, which leads us 
to the {\bf Fano form} \cite{Fa83}
\begin{equation} 
\rho = \frac{1}{NK}\Bigl[
{\mathbbm 1}_{NK} + 
\! \sum_{i=1}^{N^2-1}\tau^{A}_i \sigma_i \otimes {\mathbbm 1}_K +
\! \sum_{j=1}^{K^2-1} \tau^{B}_j   {\mathbbm 1}_N \otimes \sigma_j +
\! \sum_{i=1}^{N^2-1}\sum_{j=1}^{K^2-1} \beta_{ij} \sigma_i \otimes \sigma_j
\Bigr]  \ .
\label{Fano}
\end{equation}
\noindent Here $\vec{\tau}^{A}$ and $\vec{\tau}^{B}$
are Bloch vectors of the partially reduced states,
while a real $(N^2-1) \times (K^2-1)$ matrix $\beta$ 
describes the correlation between both subsystems.
If $\beta=0$ then the 
state is separable,
but the reverse is not true.
Note that 
for product states $M_{ij}\equiv \beta_{ij}-\tau^A_i\tau^B_j=0$,
hence the norm $||M||^2$ characterises to what extent 
$\rho$ is not a product state \cite{SM95}. 
Keeping both Bloch vectors constant 
and varying $\beta$ in such a way to preserve 
positivity of $\rho$ we obtain a $(N^2-1)(K^2-1)$
dimensional family of bi-partite mixed states, which are 
locally indistinguishable.
   
The definition of separability (\ref{mixedsep}) is implicit,
so it is in general not easy to see,
if such a decomposition exists for a given density matrix.
Separability criteria found so far 
may be divided into two disjoint classes:  
{\bf A)} sufficient and necessary, 
but not practically usable; and {\bf B)} easy to use, but only necessary
(or only sufficient). A simple, albeit amazingly 
powerful criterion was found by Peres \cite{Pe96}, who 
analyzed the action of partial transposition on an arbitrary separable state, 

\begin{equation}
\rho_{\rm sep}^{T_A} \ \equiv \ (T  \otimes {\mathbbm 1}) (\rho_{\rm sep})
= \sum_j \, q_j \, (\rho^A_j)^T \otimes \rho^B_j  \ge 0.
\label{PPT0}
\end{equation}
\index{transposition!partial}
Thus any separable state has a positive partial transpose (is PPT),
so we obtain directly

\smallskip
\noindent
{\bf B1. PPT criterion}. {\sl If} 
$\rho^{T_A} \not\geq 0$, {\sl the state $\rho$ is entangled.}
 \smallskip

\noindent
Is is extremely easy to use: all we need to do is to perform the
partial transposition of the density matrix in question,
diagonalise, and check if all eigenvalues are non--negative.
Although partial transpositions were already defined in (10.28),
let us have a look, how  both operations act on a block matrix,
\begin{equation}
X=  \left[
\begin{array}{c c}
 A & B \\
 C & D
\end{array}
\right],
\quad
X^{T_B}  \equiv \left[
\begin{array}{c c}
 A^T & B^T \\
 C^T & D^T
\end{array}
\right] \ ,
\quad
X^{T_A}  \equiv \left[
\begin{array}{c c}
 A & C \\
 B & D
\end{array}
\right] \ .
\label{partaltra2}
\end{equation}

\noindent
Note that $X^{T_B}=(X^{T_A})^T$, so the spectra of the two operators
are the same and the above criterion may be equivalently formulated
with the map $T_B =({\mathbbm 1} \otimes  T)$. Furthermore, 
partial transposition applied on a density matrix
produces the same spectrum
as the transformation of flipping one of both 
Bloch vectors present in its Fano form (\ref{Fano}).
Alternatively one may change the signs of all generators $\sigma_j$
of the corresponding group. For instance, flipping the second 
of the two subsystems of the same size we obtain  
\index{Fano form} 
\begin{equation} 
\rho^{F_B} \equiv  \frac{1}{N^2}
\Bigl[ {\mathbbm 1}_{N^2} + 
\sum_{i=1}^{N^2-1}\tau^{A}_i \sigma_i \otimes {\mathbbm 1}_N -
\sum_{j=1}^{N^2-1} \tau^{B}_j   {\mathbbm 1}_N \otimes \sigma_j -
\sum_{i,j=1}^{N^2-1} \beta_{ij} \sigma_i \otimes \sigma_j
\Bigr] \ ,
\label{FanoB}
\end{equation}
with the same spectrum as $\rho^{T_A}$ and $\rho^{T_B}$.
In the two--qubit case,
 reflection of all three components of the Bloch vector, 
${\vec \tau}^B \to  -{\vec \tau}^B$,  
is equivalent to changing the sign of its single component $\tau^B_y$ (partial
transpose),
followed by the $\pi$--rotation along the $y$--axis.

To watch the PPT criterion in action
consider the family of 
{\it generalised Werner states}\,
which interpolate between maximally mixed state $\rho_*$
and the maximally entangled state $P_+=|\phi^+\rangle \langle \phi^+|$, 
(For the original 
{\it Werner states} \cite{We89} the singlet pure state 
$|\psi^-\rangle=(|01\rangle -|10\rangle)/\sqrt{2}$
was used instead of $|\phi^+\rangle$),
\begin{equation}
\rho_W(x) = x |\phi^+\rangle \langle \phi^+|
 +(1-x) \frac{1}{N} {\mathbbm 1}
{\rm \quad with  \quad} x\in [0,1] \ .
 \label{Wernern}
\end{equation}
One eigenvalue equals $[1+(N-1)x]/N$, and the
remaining $(N-1)$ eigenvalues are degenerate and equal
to $(1-x)/N$.
In the $N=2$ case: 
{\small
\begin{equation}
\rho_x=
\frac{1}{4}
\left[
\begin{array}{c c c c}
 1+x & 0   & 0  & 2x \\
 0   & 1-x & 0  & 0 \\
 0   & 0   & 1-x & 0 \\
 2x  & 0    & 0 &  1+x
\end{array}
\right] ,
\quad
\rho_x^{T_A}=
\frac{1}{4}
\left[
\begin{array}{c c c c}
 1+x & 0   & 0  &  0 \\
 0   & 1-x & 2x  & 0 \\
 0   & 2x   & 1-x & 0 \\
 0   & 0   & 0  &  1+x
\end{array}
\right] .
\label{Wern2}
\end{equation}
}

\noindent Diagonalisation of the partially
transposed matrix $\rho_x^{T_A}=\rho_x^{T_B}$
gives the spectrum
$\frac{1}{4} \{1+x,1+x,1+x,1-3x\}$.
This matrix is positive definite if $x\le 1/3$,
hence Werner states are entangled for $x> 1/3$.
It is interesting to observe that
the critical state $\rho_{1/3}\in \partial {\cal M}_{\rm sep}$,
is localized at the distance
$r_{\rm in}=1/\sqrt{24}$ from the maximally mixed state $\rho_*$,
so it sits on the insphere, 
the maximal sphere that one can inscribe into the set ${\cal M}^{(4)}$ of 
$N=4$ mixed states. 

As we shall see below  the PPT criterion 
works in both direction only if dim$({\cal H}) \le 6$,
so there is a need for other separability 
criteria 
 \cite{LBCKKSST00,HHH00b,BCHHKLS02,Te02,Br02}.
Before reviewing the most important of them 
let us introduce one more notion often used in the physical literature.

An Hermitian operator $W$ is called an {\it entanglement witness}
for a given entangled state $\rho$ if Tr$\rho W<0$
and Tr$\rho\sigma \ge 0$ for all separable $\sigma$ \cite{HHH96a,Te00}.
For convenience the normalisation Tr$W=1$ is assumed.
Horodeccy \cite{HHH96a} proved a useful

\index{lemma!witness}  
\smallskip
\noindent
{\bf Witness lemma.}
{\sl For any entangled state $\rho$ there exists
   an entanglement witness $W$.}
\smallskip

\noindent In fact this is the Hahn-Banach separation theorem
 (section 1.1) in slight disguise. 

It is instructive to realise a direct relation with the
dual cones construction  discussed in chapter 11:
any witness operator is  proportional to a dynamical matrix,
$W=D_{\Phi}/N$, corresponding to a non-completely positive map $\Phi$.
Since $D_{\Phi}$ is block positive (positive on product states),
the condition Tr$\,W \sigma \ge 0$
holds for all separable states for which the decomposition
(\ref{mixedsep}) 
exists.
Conversely, a state $\rho$ is separable if Tr$\,W \rho\ge 0$
for all block positive $W$.
This is just the definition (11.15)
of a super-positive map $\Psi$. We arrive, therefore,
at a key observation:
the set ${\cal SP}$ of super-positive maps  is isomorphic
with the set ${\cal M}_S$ of separable states 
by the Jamio{\l}kowski isomorphism, $\rho=D_{\Psi}/N$.

An intricate link between positive maps
and the separability problem is made clear in the

\smallskip
\noindent
 {\bf A1. Positive maps criterion} \cite{HHH96a}. 
{\sl A state $\rho$
is separable if and only if  $\rho'=(\Phi \otimes {\mathbbm 1}) \rho$
is positive for all positive maps $\Phi$}.
\smallskip

\noindent
To demonstrate that this condition is necessary,  act
with an extended map on the separable state (\ref{mixedsep}),
\begin{equation}
\Bigl( \Phi \otimes {\mathbbm 1} \Bigr) \Bigl( \sum_j q_j \,  \rho^A_j \otimes
\rho^B_j
\Bigr) \: = \:  \sum_j q_j \, \Phi ( \rho^A_j)  \otimes \rho^B_j  \:  \ge \: 0
\: .
 \label{poscrit}
\end{equation}
Due to positivity of $\Phi$ the above combination of positive operators is
positive.
To prove sufficiency, assume that
$\rho'= (\Phi \otimes {\mathbbm 1})\rho$ is positive. Thus
Tr$\rho' P\ge 0$ for any projector $P$. Setting $P=P_+=|\phi^+\rangle \langle
\phi^+|$
and making use of the adjoint map we get
Tr$\rho (\Phi \otimes {\mathbbm 1})P_+=
\rm \frac{1}{N}{Tr}\rho D_{\Phi} \ge 0$. 
Since this property holds for all positive maps $\Phi$,
it implies separability of $\rho$ due to
the witness lemma $\square$.

The positive maps criterion holds also if
the map acts on the second subsystem.
However, this criterion is not easy to use:
one needs to verify that the required inequality is satisfied for {\sl all}
positive maps.  The situation becomes simple for the
$2 \times 2 $ and $2 \times 3$ systems.  
In this case any positive map is decomposable
due to the St{\o}rmer--Woronowicz theorem 
and may be written as a convex combination
of a CP map and a CcP map, which involves the transposition $T$ 
(see section 11.1). Hence, to apply the above criterion
we need to perform one check working with the partial transposition
$T_A =(T  \otimes {\mathbbm 1})$. In this way we become

\smallskip
\noindent
{\bf B1'. Peres--Horodeccy criterion} \cite{Per96,HHH96a}.
{\sl A state $\rho$ acting on ${\cal H}_2 \otimes {\cal H}_2$
(or ${\cal H}_2 \otimes {\cal H}_3$) composite Hilbert 
space is separable if and only if}
$\rho^{T_A} \ge 0$.
\smallskip

\index{states!PPT}
\index{maps!PPT--inducing}
\index{maps!PPT--preserving}
In general, the set of bi--partite states may be
divided into {\it PPT states} (positive partial transpose)
and {\it NPPT states} (not PPT).
A map $\Phi$ is related 
by the Jamio{\l}kowski isomorphism  
to a PPT state if  $\Phi \in {\cal CP}\cap {\cal C}c{\cal P}$.
Complete co--positivity of $\Phi$ implies that 
$T\Phi$ is completely positive, so 
$(T\Phi \otimes {\mathbbm 1})\rho \ge 0$
for any state $\rho$.
Thus $\rho'=(\Phi \otimes {\mathbbm 1})\rho$,
is a PPT state, 
so such  a map may be called {\it PPT--inducing},
 ${\cal PPTM}\equiv {\cal CP}\cap {\cal C}c{\cal P}$.
These
maps should not be confused with {\it PPT--preserving} maps 
\cite{Ra01,EVWW01},
which act on bi--partite systems and fulfill another property:
if $\rho^{T_A} \ge 0$ then $(\Psi(\rho))^{T_A} \ge 0$.

Similarly, a super-positive map $\Phi$ is related by the isomorphism
(11.22)  with a separable state. Hence 
$(\Phi \otimes {\mathbbm 1})$ acting 
on the maximally entangled state $|\phi^+\rangle \langle \phi^+|$ 
is separable. It is then not surprising that 
$\rho'=(\Phi \otimes {\mathbbm 1})\rho$
becomes separable for an arbitrary state $\rho$  \cite{HSR03}, 
which explains why  SP maps are also called
{\it entanglement breaking channels}.
Furthermore, due to the positive maps criterion 
 $(\Psi \otimes {\mathbbm 1})\rho' \ge 0$
for any positive map $\Psi$.
In this way we have arrived at the first of three 
duality conditions 
equivalent to (11.16-11.18),
\begin{eqnarray}
\{ \Phi \in {\cal SP} \} \ \ \Leftrightarrow \ \  & \Psi\cdot \Phi \in {\cal
CP}
 &{\rm \quad for \quad all \quad}  \Psi \in {\cal P}\ , \\ 
\{ \Phi \in {\cal CP} \}\ \ \Leftrightarrow \ \  & \Psi\cdot \Phi \in {\cal CP} 
&{\rm \quad for \quad all \quad}  \Psi \in {\cal CP} \ , \\ 
\{ \Phi \in {\cal P} \}\ \ \Leftrightarrow  \ \ & \Psi\cdot \Phi \in {\cal CP} 
&{\rm \quad for \quad all \quad}  \Psi \in {\cal SP} \ . 
 \label{dualbis}
\end{eqnarray}
\index{cones!dual}
\index{positivity!complete}
\noindent The second one reflects the fact that a composition of two 
CP maps is CP, while the third one is dual to the first.

Due to the St{\o}rmer and Woronowicz theorem and
the Peres--Horodeccy criterion, 
all PPT states for  $2 \times 2$ and $2 \times 3$
problems are separable (hence any PPT-inducing map is SP)
 while all NPPT states are
entangled. In higher dimensions there exist
PPT entangled states (PPTES), and this fact motivates
investigation of positive, non--decomposable maps and 
other separability criteria.

\smallskip
\noindent
 {\bf B2. Range criterion}  \cite{Ho97}.
{\sl If a state $\rho$ is separable,
then there exists a set of  
pure product states
such that  $|\psi_i \otimes \phi_i\rangle$ span the range of $\rho$
 and $T_B(|\psi_i \otimes \phi_i\rangle$ span the range of $\rho^{T_B}$}.

The action of the partial transposition
on a product state gives
$|\psi_i \otimes \phi_i^{*}\rangle$, where  
$^*$ denotes complex conjugation 
in the standard basis. This criterion,
proved by P. Horodecki \cite{Ho97},
allowed him to identify the first 
PPTES in the $2 \otimes 4$ system.
Entanglement of $\rho$ was detected by 
showing that none of the product states
from the range of $\rho$,
if partially conjugated, belong to the range of $\rho^{T_B}$.

The range criterion allows one to construct 
PPT entangled states related to
{\it unextendible product basis}, (UPB).
It is a set of  orthogonal product vectors $|u_i\rangle \in {\cal H}_N \otimes
{\cal H}_M$,
$i=1,\dots,k<MN$,
such that there does not exist any product vectors orthogonal to all of them 
\cite{BVMSST99,AL01,VMSST03}.
 We shall recall an example found in \cite{BVMSST99} for $3 \times 3$ system,
\begin{eqnarray}
|u_1\rangle &=& \frac{1}{\sqrt{2}} |0\rangle \otimes |0-1\rangle , \quad 
|u_2\rangle = \frac{1}{\sqrt{2}} |2\rangle \otimes |1-2\rangle , \quad 
|u_3\rangle = \frac{1}{\sqrt{2}} |0-1\rangle \otimes |2\rangle , \nonumber \\  
|u_4\rangle &=& \frac{1}{\sqrt{2}} |1-2\rangle \otimes |0\rangle , \quad 
|u_5\rangle = \frac{1}{3} |0+1+2\rangle \otimes |0+1+2\rangle . 
 \label{UPB1}
\end{eqnarray}
These five states are mutually orthogonal. However,
since they span full three dimensional spaces
in both subsystems, no product state may be 
orthogonal to all of them.

For a given UPB let $P=\sum_{i=1}^k |u_i\rangle  \langle u_i|$
denote the projector on the space  spanned by these product vectors.
Consider the mixed state, uniformly covering the complementary 
subspace, 
\begin{equation}
\rho\ \equiv \ \frac{1}{MN-k}\ ({\mathbbm 1}-P)  \ .
\label{UPB2}
\end{equation}
By construction this subspace does not contain any product
vectors, so  $\rho$ is entangled due to the range criterion.
On the other hand, the projectors $(|u_i\rangle  \langle u_i|)^{T_B}$
are mutually orthogonal, so the operator
$P^{T_B}=\sum_{i=1}^k (|u_i\rangle  \langle u_i|)^{T_B}$
is a projector. So is $({\mathbbm 1}-P)^{T_B}$,
hence $\rho^{T_B}$ is positive.
Thus the state (\ref{UPB2}) is a positive partial transpose entangled 
state.
The UPB method was used to construct PPTES
in \cite{BVMSST99,BP00,VMSST03,Pi04}, while  
not completely positive maps 
were applied in \cite{HKP03,BFP04} for this purpose.
Conversely, PPTES were used in  \cite{Te00}
to find non-decomposable positive maps.

\medskip
\noindent
{\bf  B3. Reduction criterion}  \cite{CAG99,HH99}. {\sl If a state $\rho$
is separable then the reduced states $\rho_A ={\rm Tr}_B \rho$
and $\rho_B ={\rm Tr}_A \rho$ satisfy}
\begin{equation}
\rho_A \otimes {\mathbbm 1} -\rho \ge 0
{\quad \rm and \quad}
 {\mathbbm 1}\otimes \rho_B -\rho \ge 0  \ .
\label{red1}
\end{equation}

This statement follows directly from the
positive maps criterion with the map
 $\Phi(\sigma)=({\rm Tr} \sigma){\mathbbm 1}-\sigma$
applied to the first or the second subsystem.
Computing the dynamical matrix for this
map composed with the transposition, $\Phi'=\Phi T$,
we find that $D_{\Phi'}\ge 0$, hence $\Phi$ is CcP
and (trivially) decomposable. Thus the reduction
criterion cannot be stronger than the PPT criterion
 (which is the case for the
{\sl generalised reduction criterion} \cite{ACF03}).

There exists, however, a good
reason to pay some attention to the reduction criterion:
the Horodeccy brothers have shown \cite{HH99} that any state $\rho$
violating (\ref{red1}) is {\it distillable},
i.e. there exists a LOCC protocol which allows one to
extract locally  maximally entangled states
out of $\rho$ or its copies \cite{BVSW96,Ra99b}.
Entangled states, which are not distillable are
called {\it bound entangled} \cite{HHH98,HHH99}.

A general question, which mixed state may be distilled
is not solved yet \cite{BCHHKLS02}.
(Following
literature we use two similar terms:
entanglement {\sl concentration} and {\sl distillation},
for local operations performed on pure and mixed states, respectively.
While the former operations are reversible, the latter are not.)
Again the situation is clear for systems
with dim$({\cal H})\le 6$: all PPT states are
separable, and all NPPT states are entangled and distillable.
For larger systems there exists 
PPT entangled states
and all of them are not distillable, hence bound entangled \cite{HHH98}.
(Interestingly, there are no bound entangled
states of rank one nor two \cite{HSTT03}.)
 Conversely, one could think that
all NPPT entangled states are distillable, but this seems
not to be the case \cite{DCLB00,VSSTT00}.

\index{majorisation} 
\smallskip
\noindent
\index{majorisation}
{\bf B4. Majorisation criterion}  \cite{NK01}.
{\sl If a state $\rho$ is separable,
then the  reduced states $\rho_A$ and $\rho_B$
satisfy the majorisation relations
}
\begin{equation}
\rho \prec \rho_{A}
{\rm \quad \quad and  \quad \quad}
\rho \prec  \rho_{B} \ .
 \label{sepmajor}
\end{equation}

\noindent
In brief, separable states are more disordered globally than locally.
To prove this criterion one needs to find a bistochastic matrix $B$
such that the spectra satisfy ${\vec \lambda} = B {\vec \lambda_A}$.
The majorisation relation implies that any Schur convex functions satisfies
the inequality (2.8). For Schur concave functions the direction of 
the inequality changes. In particular, the

\smallskip
\noindent
{\bf B5. Entropy criterion}.
{\sl If a state $\rho$ is separable,
then the R{\'e}nyi entropies fulfill}
\begin{equation}
S_{q}(\rho) \ge S_{q}(\rho_A)
{\rm \quad and  \quad }
S_{q}(\rho) \ge S_{q}(\rho_B)
{\quad \rm for \quad } q\ge 0\ ,
 \label{sepentren}
\end{equation}
\noindent
follows.
 The entropy criterion was originally formulated
for $q=1$ \cite{HH94}. Then this statement may be
equivalently expressed in terms of the
{\it conditional entropy},
$S(A|B)=S(\rho_{AB})- S(\rho_A)$:
for any separable bi-partite  state $S(A|B)$ is non--negative.
(The opposite quantity, 
 $-S(A|B)$, is called {\it coherent quantum information} \cite{SN96}
and plays an important role in quantum communication \cite{HHHO05}.)
 
Thus negative conditional entropy 
of a state $\rho_{AB}$ confirms its entanglement \cite{HH96,SN96,CA99}.
The entropy criterion was proved for $q=2$ in \cite{HHH96b} 
and later formulated also for the Havrda--Charvat--Tsallis
entropy (2.77) \cite{AR01,TLB01,RR02,RC02}.
Its combination with the entropic uncertainty relations
of Massen and Uffink \cite{MU88}
provides yet another interesting family of separability criteria \cite{GL04}.
However, it is worth to emphasize that in general the spectral 
properties do not  determine separability---there exist pairs of isospectral 
states, one of which is separable, the other not.

\smallskip
\noindent
\index{contraction}
{\bf A2. Contraction criterion}. {\sl  A bi-partite state $\rho$
is separable if and only if any extended trace preserving positive map
act as a (weak) contraction in sense of the trace norm,}
\begin{equation}
||\rho'||_{\rm Tr} \: =\: 
||({\mathbbm 1} \otimes \Phi)\rho||_{\rm Tr}  \: \le \:   
||\rho||_{\rm Tr} \: = \: {\rm Tr} \rho \:  = \:  1 \ .
 \label{sepcontr}
\end{equation}
\smallskip

This criterion was formulated in \cite{HHH02}
basing on earlier papers \cite{Rud02,CW03}.
To prove it notice that the sufficiency follows
from the positive map criterion: since Tr$\rho'=1$,
hence $||\rho'||_{\rm Tr}\le 1$ implies that $\rho' \ge 0$.
To show the converse consider a normalised product state 
$\rho=\rho^A \otimes \rho^B$.
Any trace preserving  positive map $\Phi$
acts as isometry in sense of the trace norm,
and the same is true for the extended map,
\begin{equation}
||\rho'||_{\rm Tr} = 
\bigl| \bigr| ({\mathbbm 1}\otimes \Phi)(\rho^A \otimes \rho^B)
\bigl|\bigr|_{\rm Tr}=
||\rho^A|| \cdot ||\Phi(\rho^B)||_{\rm Tr}  =  1 \ .
 \label{sepcontr2}
\end{equation}
\smallskip
Since the trace norm is convex, $||A+B||_{\rm Tr} \le ||A||_{\rm Tr}+||B||_{\rm
Tr}$,
any separable state fulfills 
\begin{equation}
\Bigl| \Bigr|  ({\mathbbm 1}\otimes \Phi) 
\Bigl( \sum_i q_i (\rho^A_i \otimes \rho^B_i) \Bigr)\Bigl| \Bigr|_{\rm Tr}
\ \le \  \sum_i q_i \,|| \rho^A_i \otimes \Phi(\rho^B_i ) ||_{\rm Tr} 
\ =\ \sum_i q_i =1 \ ,
 \label{sepcontr3}
\end{equation}
which ends the proof. $\square$.

Several particular cases of this criterion
could be useful. Note that the celebrated PPT criterion {\bf B1}
follows directly, if the transposition $T$ is selected
as a trace preserving map  $\Phi$, since the norm condition,
$||\rho^{T_A}||_{\rm Tr} \le 1$, implies  positivity, $\rho^{T_A} \ge 0$.
Moreover, one may formulate an analogous criterion
for global maps $\Psi$, which act as contractions
on any bi--partite product states, 
$||\Psi(\rho_A \otimes \rho_B)||_{\rm Tr} \le 1$. 
As a representative example let us mention

\smallskip
\noindent
\index{reshuffling}
 {\bf B6. Reshuffling criterion}.
(also called also {\it realignement} criterion \cite{CW03}
or {\it computable cross-norm criterion} \cite{Rud03}).
{\sl  If a bi-partite state $\rho$
is separable then reshuffling (10.33)
does not increase its trace norm,}
\begin{equation}
||\rho^R||_{\rm Tr} \ \le \  ||\rho||_{\rm Tr}=1 \ .
 \label{reshcontr}
\end{equation} 
\smallskip
We shall start the proof considering an
 arbitrary product state, $\sigma_A \otimes \sigma_B$.
By construction its Schmidt decomposition consists of one term only.
This implies
\begin{equation}
||(\sigma_A \otimes \sigma_B )^R||_{\rm Tr} = 2\, ||\sigma_A||_2 \cdot
||\sigma_B||_2=
\sqrt{{\rm Tr} \,\sigma_A^2}\, \sqrt {{\rm Tr}\, \sigma_B^2} \ \le \ 1 \ . 
\label{reshprod}
\end{equation}
Since the reshuffling transformation is linear, 
$(A+B)^R = A^R+B^R$,
and the trace norm is convex, 
any separable state satisfies
\begin{equation}
\Bigl| \Bigr|\Bigl( \sum_i q_i (\sigma^A_i \otimes \sigma^B_i) \Bigr)^R
\Bigl|\Bigr|_{\rm Tr} 
\ \le \
 \sum_i q_i || (\sigma^A_i \otimes \sigma^B_i )^R ||_{\rm Tr}
\ \le \ \sum_i q_i =1 \ ,
 \label{reshcont2}
\end{equation}
which completes the reasoning. $\square$.

In the simplest case of two qubits,  
the latter criterion is weaker than the PPT: 
examples of NPPT states, the entanglement of which is 
not detected by reshuffling, were provided by Rudolph \cite{Rud03}.
However, for some larger dimensional problems
the reshuffling criterion becomes useful, since it is capable of detecting
PPT entangled states, for which  $||\rho^R||_{\rm Tr} > 1$ 
\cite{CW03}.

There exists several
other separability criteria, not discussed here. 
Let us mention applications of the range criterion
for $2 \times N$ systems \cite{DCLB00},
checks for low rank density matrices \cite{HLVC00},
reduction of the dimensionality of the problem \cite{Woe04},
relation between purity of a state and its
maximal projection on a pure states \cite{LBCKKSST00},
or criterion obtained by expanding a mixed state
in the Fourier basis \cite{PR00}.
The problem which separability criterion
 is the strongest, and what 
the implication chains among them are, remains a subject of 
a vivid research \cite{VW02,ACF03,CW04,BPCP04}.
In general, the separability problem is 'hard',
since it is known that it belongs to the
NP complexity class \cite{Gu03}.
Due to this intriguing mathematical result
it is not surprising that
all operationally feasible analytic 
criteria provide  partial solutions only.
On the other hand, one should appre\-cia\-te practical methods
constructed to decide separability numerically.  
Iterative algorithms based on an extension of the PPT criterion 
for higher dimensional spaces \cite{DPS02,DPS04}
or non-convex optimization \cite{EHGC04} 
are able to detect 
the entanglement in a finite number of steps.
Another  algorithm provides an explicit decomposition 
into pure product states \cite{HBr04},
confirming that the given mixed state $\rho$
is separable. A combination of these two  approaches
 terminates after a finite time $t$
and gives an inconclusive answer only 
if $\rho$ belongs to the $\epsilon$--vicinity of the boundary of 
the set of separable states. By increasing the computation time $t$
one may make the width $\epsilon$ of the 'no--man's land' 
arbitrarily small.

\section{Geometry of the set of separable states}
\label{sec:geomsep}

Equipped with a broad spectrum of separability criteria, 
we may try to describe the structure of the set 
${\cal M}_S$ of the separable states. This task becomes easier  
for the two--qubit system, for which 
positive partial transpose implies separability.
Hence the set of $N=4$ separable states arises
as an intersection of the entire body of mixed states with its 
reflection induced by partial transpose,
\begin{equation}
 {\cal M}^{(4)}_S={\cal M}^{(4)}\ \cap \ T_A({\cal M}^{(4)}) \ 
\label{separb4}
\end{equation}
This observation suggests that the 
set of separable states has a positive volume.
The maximally mixed state is invariant with respect to partial transpose, 
$\rho_*=\rho_*^{T_B}$ and occupies the center of the body ${\cal M}^{(4)}$.
It is thus natural to ask, what is the radius of the 
separable ball centered at $\rho_*$?
The answer its very appealing in the simplest, Euclidean geometry:
the entire maximal $15$-D ball inscribed in ${\cal M}^{(4)}$ is separable 
\cite{ZHSL98}. Working with the distance $D_2$ defined in eq. ({9.26}), 
its radius reads $r_{\rm in}=1/\sqrt{24}$.

\begin{figure} [htbp]
   \begin{center} \
 \includegraphics[width=10.5cm,angle=0]{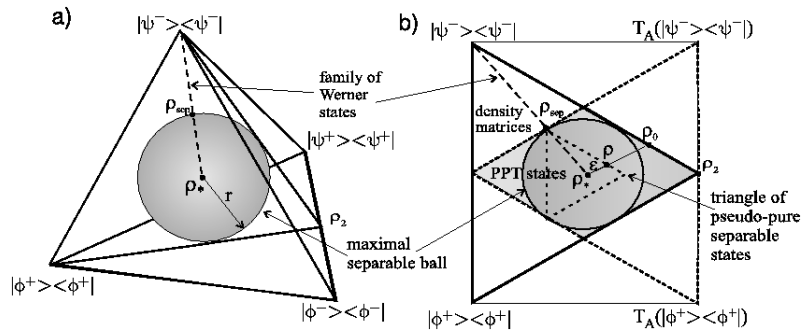}
\caption{Maximal ball inscribed inside the $15$--D body ${\cal M}^{(4)}$
of mixed states
is separable: a) $3$--D cross-section containing four Bell states,
 b)  $2$--D cross--section defined by two Bell states and $\rho_*$
with the maximal separable triangle of pseudo-pure states.} 
\label{fig:ent04}
\end{center}
 \end{figure}

The separable ball is sketched in two or three dimensional
cross--sections of ${\cal M}^{(4)}$ in Fig. \ref{fig:ent04}. To prove its 
separability we shall invoke  

\smallskip
\noindent
{\bf Mehta's Lemma} \cite{Me89}.
{\sl Let $A$ be a Hermitian matrix of size $D$
and let $\alpha={\rm Tr} A/\sqrt{{\rm Tr} A^2}$.
If $\alpha\ge \sqrt{D-1}$ then $A$ is positive.}
\smallskip

Its proof begins with an observation
that both traces are basis independent, 
so we may work in the eigenbasis of $A$.
Let $(x_1,\dots x_D)$ denote the spectrum of $A$.
Assume first that one eigenvalue, say $x_1$,  is negative.
Making use of the right hand side of the standard estimation between 
the ${l}_1$ and ${l}_2$--norms (with prefactor $1$) of an $N$--vector,
 $||A||_2 \le ||A||_1 \le N ||A||_2$,
we infer
\begin{equation}
 {\rm Tr A}=\sum_{i=1}^D x_i < \sum_{i=2}^D x_i
\le \sqrt{D-1} \Bigl( \sum_{i=2}^D x_i^2 \Bigr)^{1/2} <
\sqrt{D-1} \sqrt{{\rm Tr} A^2} \ .
\label{mehta}
\end{equation}
This implies that $\alpha < \sqrt{D-1}$.
Hence if the opposite is true and $\alpha \ge \sqrt{D-1}$ then none of
the eigenvalues $x_i$ could be
negative, so $A\ge 0$. $\square$.

\index{transposition!partial}  
The partial transpose preserves the trace and 
the HS norm of any state,
$||\rho^{T_B}||^2_2=||\rho||^2_2=\frac{1}{2}\, {\rm Tr}\rho^2$.
Taking for $A$ a partially transposed density matrix $\rho^{T_B}$ 
we see that $\alpha^2=1/{\rm Tr}\rho^2$.
Let us apply the Mehta lemma to an arbitrary mixed state  
of a $N \times N$ bipartite system, for which
the dimension $D=N^2$,
\begin{equation}
1/{\rm Tr}\rho^2 \ \ge\  N^2-1 \quad \Rightarrow
\quad \rho \quad
{\rm is \quad PPT. \quad }
\label{pptn}
\end{equation}

Since the purity condition Tr$\rho^2=1/(D-1)$ characterizes the 
insphere of 
${\cal M}^{(D)}$,
we conclude that for any bipartite
(or multipartite \cite{KZM02})
system the entire maximal ball
inscribed inside the set of mixed states 
consists of PPT states only. 
This property implies separability 
for $2 \times 2$ systems.
An explicit separability decomposition (\ref{mixedsep}) 
for any state inside the ball was provided in \cite{BCJLPS99}.
Separability of the maximal ball
for higher dimensions 
was established by Gurvits and Barnum \cite{GB02},
who later estimated the radius of the separable
 ball for multipartite systems \cite{GB03,GB04}.

For any $N \times N$ system the volume 
of the maximal separable ball, ${\bf B}^{\rm sep}_{N^4-1}$ 
may be compared with the 
Euclidean volume  of ${\cal M}^{(N^2)}$.
The ratio 
\begin{equation}
\frac{{\rm Vol} ({\bf B}^{\rm sep}_{N^4-1})}
      { {\rm Vol}\bigl( {\cal M}^{(N^2)} \bigr)} = 
\frac{\pi^{(N^2-1)/2} \, 2^{(N^2-N^4)/2}\, \Gamma(N^4) }
     { \Gamma [(N^4+1)/2]\ N^{N^4} \, (N^2-1)^{(N^4-1)/2}\
 \prod_{k=1}^{N^2} \Gamma (k)}
\label{volrat}
\end{equation}
decreases fast with $N$, which suggests
that for higher dimensional systems
the separable states are not typical.
The actual probability $p$ to find
a separable mixed state is positive for any finite $N$
and depends on the measure used \cite{Zy99,Sl99,Sl04}.
However, in the limit $N \to \infty$ the set of
separable states is nowhere dense \cite{CH00},
so the probability $p$ computed with respect to an
arbitrary non--singular measure tends to zero.

Another method of exploring the vicinity of the 
maximally mixed state consists in studying 
{\it pseudo--pure states} 
\begin{equation}
\rho_{\epsilon} \equiv \frac{\mathbbm 1}{N^2}
 (1-\epsilon) +\epsilon |\phi\rangle \langle \phi | \ ,
\label{pseudopure}
\end{equation}
which are relevant for experiments with
nuclear magnetic resonance (NMR)
for $\epsilon \ll 1$.
The set ${\cal M}_{\epsilon}$ is then defined 
as the convex hull of all $\epsilon$--pseudo pure states.
It forms a smaller copy of the entire set  
of mixed states of the same shape 
and is centered at $\rho_*={\mathbbm 1}/{N^2}$.
\index{states!maximally mixed}  

For instance, since the cross-section of the set ${\cal M}^{(4)}$
shown in Fig. \ref{fig:ent04}b is a triangle,
so is the set ${\cal M}_{\epsilon_c}$--a dashed triangle
located inside the dark rhombus of separable states.
The rhombus is obtained as a cross-section of the 
separable octahedron,
 which arises as a common part of the tetrahedron of density matrices
spanned by four Bell states 
and its reflection representing their partial transposition \cite{HH96,Ar97}.
An identical octahedron of super-positive maps 
will be formed by intersecting 
the  tetrahedrons of  CP and CcP
one--qubit unital maps shown  in Fig. 10.4.
Properties of a separable octangula obtained
 for other 3-D cross-sections of ${\cal M}^{(4)}$
 were analysed in \cite{Er02}. 
Several 2-D cross-sections plotted in \cite{JS01,VDM02}
provide further insight into
the geometry of the problem.

Making use of the radius (\ref{pptn}) of the separable ball
we obtain that the states ${\cal M}_{\epsilon}$ 
of a $N \times N$ bipartite
system are separable for $\epsilon \le \epsilon_c=1/(N^2-1)$.
 Bounds for $\epsilon_c$
in multipartite systems were 
 obtained in \cite{BCJLPS99,DMK00,PR02,GB03,Sza04,GB04}.
The size of the separable ball is large enough
that to generate a genuinely entangled pseudo-pure state
in an NMR experiment one would need to deal with at least $34$ qubits
\cite{GB04}.
Although experimentalist gained full control 
over $7-10$ qubits up till now, and work with separable states only,
the NMR quantum computing does fine \cite{CFH97,CGKL98,LCNV02}.

Usually one considers states separable 
with respect to a given decomposition of the 
composed Hilbert space, 
${\cal H}_{N^2}={\cal H}_{A}\otimes {\cal H}_{B}$.
A state $\rho$ may be separable with respect to 
a given decomposition and entangled with respect to another one.
Consider for instance, two decompositions
of  ${\cal H}_6$: 
${\cal H}_2 \otimes {\cal H}_3$ and 
${\cal H}_3 \otimes {\cal H}_2$
which describe different physical problems.
There exist states separable with respect to 
the former decomposition and entangled 
with respect to the latter one.
On the other hand one may ask, which states
are separable with respect to all possible  
splittings of the composed system into subsystems $A$ and $B$.
This is the case if $\rho'=U\rho\, U^{\dagger}$ is separable for 
any global unitary $U$, and states possessing this property are 
called {\it absolutely separable} \cite{KZ01}.

All states belonging to the maximal ball
inscribed into the set of mixed states for a bi-partite problem
are not only separable but also absolutely separable.
In the two--qubit case the set of absolutely separable states
is larger than the maximal ball: As conjectured in \cite{IH00}
and proved in \cite{VAM01}
it contains any mixed state $\rho$ for which
\begin{equation}
C^*({\vec x}) \ \equiv \ x_1-x_3 - 2\sqrt{ x_2 x_4} \le 0 \ ,
\label{abssep4}
\end{equation}
where ${\vec x}=\{ x_1\ge x_2\ge x_3 \ge x_4\}$
denotes the ordered spectrum of $\rho$.
The problem, whether 
there exist absolutely separable states outside the maximal ball
was solved for $2 \times 3$ case \cite{Hi05},
but it remains open in higher dimensions.
Numerical investigations suggest that in such a case 
the set ${\cal M}_S$ of separable states, located in central parts of ${\cal
M}$,
is covered by a shell of bound entangled states. However
this shell is not perfect, in the sense that
the set of NPPT entangled states (occupying certain 'corners' of ${\cal M}$)  
has a common border with the set of separable states.

Some insight into the geometry of the
problem may be gained by studying the manifold 
$\Sigma$ of mixed products states. 
To verify whether a given state $\rho$
belongs to $\Sigma$ one computes the partial traces
and checks if $\rho_A \otimes \rho_B$ is equal to $\rho$.
This is the case e.g. for the maximally mixed state,
$\rho_* \in \Sigma$. All states tangent to $\Sigma$ 
at $\rho_*$ are separable, while the normal subspace contains the 
maximally entangled states. Furthermore, for any bi-partite systems
the maximally mixed state $\rho_*$ is the product state closest to 
any maximally entangled state (with respect to the HS distance) 
\cite{LSG02}. 

Let us return to characterisation of the boundary
of the set of separable states for a bi-partite system.
For any entangled state $\sigma_{\rm ent}$ one may define
the separable state  $\sigma_{\rm sep}\in \partial{\cal M}_S$,
 which is closest 
to $\sigma_{\rm ent}$ with respect to a given metric.
In general it is not easy to find the closest
separable state, even in the two qubit case, for which the
$14$-dim boundary of the set  ${\cal M}_S^{(4)}$
may be characterised explicitly, 
\begin{equation}
(\rho \in \partial {\cal M}_S^{(4)}) \ \Rightarrow \ {\rm det}\rho =0 
{\quad \rm or \quad}  {\rm det}\rho^{T_A} =0  \ .
\label{bound4}
\end{equation}

\begin{figure} [htbp]
   \begin{center} \
 \includegraphics[width=11cm,angle=0]{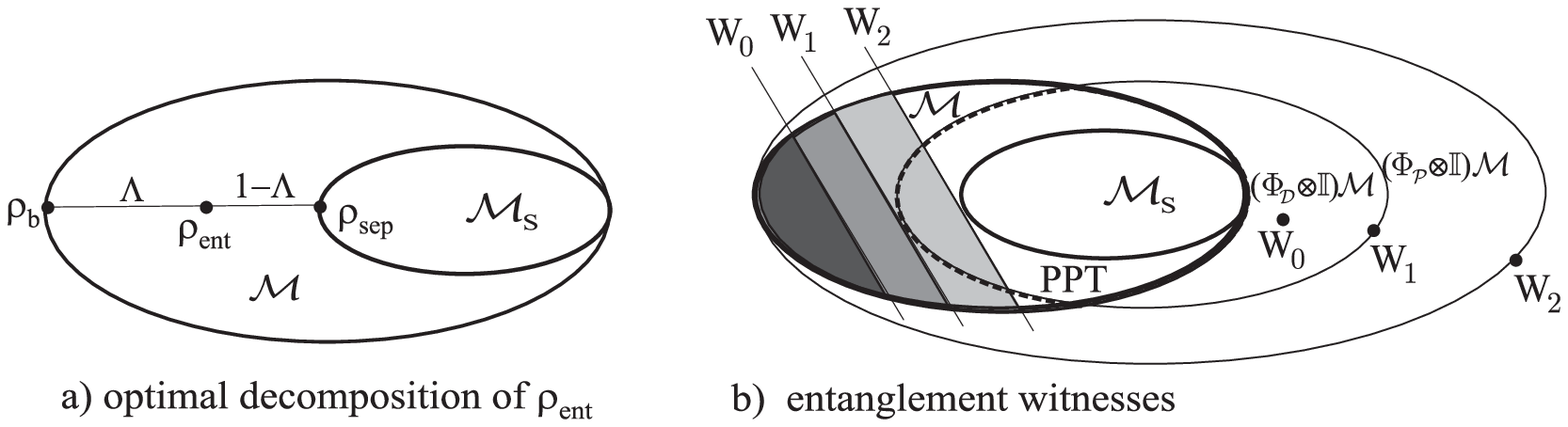}
\caption{ 
a) Best separable approximation of entangled state $\rho$;
b) a witness $W_0$ detects entanglement 
in a subset of entanglement states; $W_1$ -- optimal decomposable witness;
$W_2$ -- optimal non--decomposable witness.} 
\label{fig:ent05}
\end{center}
 \end{figure}

Alternatively, for any entangled state one defines
the {\it best separable approximation}  also called
{\it Lewenstein--Sanpera decomposition}  \cite{LS98},
\begin{equation}
\rho_{\rm ent} \ =\  \Lambda \rho_{\rm sep} + (1-\Lambda) \rho_b ,
\label{LSdecomp}
\end{equation}
where the separable state $\rho_{sep}$
and the state $\rho_b$ are chosen in such a way
that the positive weight $\Lambda \in [0,1]$ is maximal.
Uniqueness of such a decomposition was proved in \cite{LS98}
for two qubits, and in \cite{KL01}
for any bi--partite system.
In the two--qubit problem the state $\rho_{b}$ is pure,
and is maximally entangled for any full rank state $\rho$ \cite{KL01}.
An explicit form of the decomposition (\ref{LSdecomp})
was found in \cite{WK01} for a  generic two--qubit state
and  in \cite{AJ04} for some particular cases in higher dimensions.
Note the key difference in both approaches:
looking for the separable state closest  to $\rho$
we probe the boundary $\partial {\cal M}^{(N)}_S$
of the set of separable states only,
while looking for its best separable approximation
we must also take into account the boundary of the 
entire set of density matrices -- compare  Figs. \ref{fig:ent05}a.
and \ref{fig:ent06}a.

The structure of the set of separable states
may also be analyzed with use of the  
entanglement witnesses \cite{PR03},
 already defined in the previous section.
Any witness $W$, being a non-positive operator,
may be represented as a point  located far outside
the set  ${\cal M}$ of density matrices,
in its image with respect to an extended positive map,
$(\Phi_{\cal P} \otimes {\mathbbm 1})$,
 or as a line
perpendicular to the axis $OW$ 
which crosses ${\cal M}$. The states outside this line
satisfy Tr$\rho W<0$, hence their entanglement 
is detected by $W$.
A witness $W_1$ is called {\it finer}
than $W_0$ if every entangled state detected by 
$W_0$ is also detected by $W_1$.
A witness $W_2$ is called {\it optimal} if the corresponding map 
belongs to the boundary of the set of positive operators,
so the line representing $W_2$ touches the boundary of
the set ${\cal M}_S$ of separable states.
A witness related to a generic non CP map $\Phi_{\cal P} \in {\cal P}$
may be {\it optimized} 
by sending it toward the boundary of ${\cal P}$ \cite{LKCHC00}.
If a positive map $\Phi_{\cal D}$ is decomposable,
the corresponding witness, $W=D_{\Phi_{\cal D}}/N$ is called {\it
decomposable}.
Any decomposable witness cannot detect PPT bound entangled states
- see Fig. \ref{fig:ent05}b.

One might argue that in general a witness $W=D_{\Phi}/N$ is theoretically
less useful than the corresponding map $\Phi$, since the criterion  Tr$\rho
W<0$ 
is not as powerful as $N(W^R \rho^R)^R=(\Phi \otimes {\mathbbm 1}) \rho  
\ge 0$. 
However, a non--CP map  $\Phi$  cannot be realised
in nature, while an observable $W$ may be measured. 
Suitable witness operators were actually used to  detect 
quantum entanglement experimentally  
in bi--partite \cite{BMNMAM03,GHDELMS03} and 
multi--partite systems \cite{BEKWGHBLS04}.
Furthermore, the {\it Bell inequalities}  may be viewed as a kind of 
separability criterion, related to a particular entanglement witness
 \cite{Te00b,HHH00b,HGBL05},
so evidence of their violation
for certain states \cite{ADR82}
might be regarded as  an experimental detection of quantum entanglement.

\section{Entanglement measures}
\label{sec:entmeas}

\index{entanglement!measures}
We have already learned that the degree of entanglement of 
any pure state of a $N \times K$ system may be 
characterised by the entanglement entropy (\ref{entanentr})
or any other Schur concave function $f$ of the  
Schmidt vector $\vec \lambda$.
The problem of quantifying entanglement for
mixed states becomes complicated \cite{VPRK97,DHR02,HoM01}.

Let us first discuss the properties that 
any potential measure $E(\rho)$ should satisfy. Even in this 
respect experts seem not to share exactly the same opinions
\cite{BVSW96,PR97,VP98,Vi00,HHH00}.

There are three basic axioms,

\noindent
{\bf (E1) Discriminance}.\
 {\sl $E(\rho) = 0$ if and only if $\rho$ is separable,}

\noindent
{\bf (E2)  Monotonicity} (\ref{monot})
{\sl under {\bf probabilistic} LOCC},

\noindent
{\bf (E3) Convexity},
 \ {\sl $E\bigl( a\rho +(1-a)\sigma\bigr)
  \le a E(\rho)+(1-a)E(\sigma)$,
with $a \in [0,1]$.}

\noindent 
Then there are certain additional requirements,

\noindent
\hangafter=1 \hangindent=1.1cm
{\bf (E4)  Asymptotic continuity} (We follow 
\cite{HoM01} here; 
slightly different fomulations 
of this property are used in \cite{HHH00,DHR02}), 
{\sl  Let $\rho_m$ and $\sigma_m$ denote sequences of states
acting on $m$ copies of the composite Hilbert 
space, $({\cal H}_N\otimes {\cal H}_K)^{\otimes m}$}. 

\vskip -0.45cm
\begin{equation}
{\rm If \quad}
 \lim_{m\to\infty}
||\rho_m -\sigma_m||_{1} =0 
{\quad \rm then \quad} \lim_{m\to\infty}
\frac{E(\rho_m)-E(\sigma_m)}{m \ln NK} = 0 \ ,
\label{Ascont}
\end{equation}

\noindent
{\bf (E5) Additivity}. {\sl $E(\rho \otimes \sigma)\ =\ E(\rho)+E(\sigma)$
 for any $\rho, \sigma \in {\cal M}_{NK}$,}

\noindent
{\bf (E6) Normalisation}.\
 $E(|\psi^-\rangle \langle \psi^-|)=1$,

\noindent
{\bf (E7) Computability}. \
{\sl There exists an efficient method to compute $E$ for any $\rho$.}

\noindent 
There are also alternative forms of properties {\bf (E1-E5)}, \

\noindent
{\bf (E1a) Weak discriminance}. \ 
{\sl If $\rho$ is separable then $E(\rho) = 0$,}

\noindent
{\bf (E2a) Monotonicity}  {\sl under {\bf deterministic} LOCC, \ 
           $E(\rho) \ge  E[\Phi_{\rm LOCC}(\rho)]$,} 

\noindent
{\bf (E3a) Pure states convexity}. \  $E(\rho) \le \sum_i p_i E(\phi_i)$
{\it where} $\rho= \sum_i p_i |\phi_i\rangle \langle \phi_i|$,

\noindent
{\bf (E4a)  Continuity}. \
{\sl If} $||\rho -\sigma||_{\rm 1} \to 0 $ 
{\sl then}  
$|E(\rho)-E(\sigma)| \to 0$. 

\noindent
{\bf (E5a) Extensivity}. \ $E(\rho^{\otimes n})=n E(\rho)$.

\noindent
{\bf (E5b) Subadditivity}. \ $E(\rho \otimes \sigma)\ \le \ E(\rho)+E(\sigma)$.

\noindent
{\bf (E5c) Superadditivity}. \ $E(\rho \otimes \sigma)\ \ge \
E(\rho)+E(\sigma)$.

\smallskip
The above list of postulates deserves a few comments.
The rather natural 'if and only if' condition in {\bf (E1)}
is very strong: it cannot be satisfied by 
any measure quantifying the distillable entanglement, 
due to the existence of bound entangled states.
Hence one often requires the weaker property {\bf (E1a)} instead.

Monotonicity {\bf (E2)} under 
probabilistic transformations  
is stronger than monotonicity {\bf (E2a)} under deterministic LOCC.
Since local unitary operations are reversible, 
the latter property implies 

\noindent
{\bf (E2b) Invariance with respect to local unitary operations},
\begin{equation}
E(\rho)=E(U_A \otimes U_B \, \rho \,
         U_A^{\dagger} \otimes U_B^{\dagger}) \ .
\label{locinvar}
\end{equation}
Convexity property {\bf (E3)} guarantees that one cannot
increase entanglement by mixing.
Following Vidal \cite{Vi00}, we will call 
any quantity satisfying
{\bf (E2)} and {\bf (E3)}
an {\it entanglement monotone} (Some authors
require also continuity {\bf (E4a)}).
These fundamental postulates reflect the key idea  
that quantum entanglement cannot be created locally.
Or in more economical terms:
it is not possible to get any entanglement for free --
one needs to invest resources for certain 
global operations.

The postulate that any two neighbouring states
should be characterised by similar entanglement is
made precise in {\bf (E4)}. 
Let us recall here the Fannes continuity lemma
(13.36), which estimates the difference between 
von Neumann entropies of two neighbouring mixed states.
Similar bounds may also be obtained for any other R{\'e}nyi
entropy with $q>0$, but then the bounds for $S_q$ are weaker
then for $S_1$. Although $S_q$ are continuous for $q>0$,
in the asymptotic limit $n \to \infty$
only $S_1$ remains a continuous function of the state $\rho^{\otimes n}$.
In the same way the asymptotic continuity 
distinguishes the entanglement entropy based on $S_1$
from other entropy measures related to the generalised entropies $S_q$ 
\cite{Vi00,DHR02}. 

Additivity  {\bf (E5)} is a most welcome property of an optimal
entanglement measure. 
For certain measures one can show sub- or super--additivity;
additivity requires both of them.
Unfortunately this is extremely difficult to prove 
for two arbitrary density matrices, so 
some authors suggest to require extensivity {\bf (E5a)}.
Even this property is not easy to demonstrate.
However, for any measure $E$ one may consider the quantity 
\begin{equation}
E_R(\rho) \ \equiv \ \lim_{n \to \infty} \frac{1}{n} E(\rho^{\otimes n}) \ .
\label{regulmeas}
\end{equation}
If such a limit exists, then the {\it regularised} measure $E_R$
defined in this way satisfies {\bf (E5a)} by construction.
The normalisation property {\bf (E6)},
useful to compare different quantities, 
can be achieved by a trivial rescaling.

The complete wish list {\bf (E1-E7)} is very demanding, so it is not surprising
that instead of one ideal measure of entanglement fulfilling all 
required properties,  the literature contains a plethora of measures
\cite{VP98,HoM01,Br02}, each of them satisfying some axioms only...
The pragmatic wish {\bf (E7)} is an especially tough one---since we have 
learned that even the problem of deciding the separability
is a 'hard one' \cite{Gu03,Gu04}, the quantifying of entanglement cannot be
easier.
Instead of waiting for the discovery of a single, universal measure
of entanglement, we have thus no choice but to review some approaches to the
problem.
In the spirit of this book we commence with 

\smallskip
{\centerline {\bf  I. Geometric measures}}

\begin{figure} [htbp]
  \begin{center} \
 \includegraphics[width=12cm,angle=0]{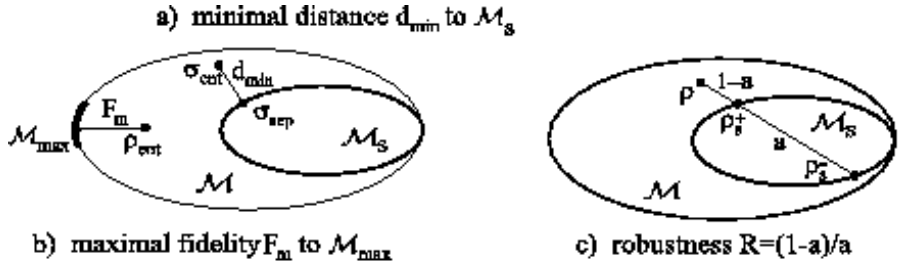}
\caption{ 
a) Minimal distance $D_x$ to the closest separable state. 
b) Maximal fidelity to a maximally entangled state. 
c) Robustness, i.e. the minimal ratio of the distance
to set ${\cal M}_S$ by its width.} 
\label{fig:ent06}
\end{center}
 \end{figure}

The distance from an analysed state $\rho$ to the set ${\cal M}_S$ of separable
states
satisfies  {\bf (E1)} by construction -- see Fig. \ref{fig:ent06}a.
However, it is not simple to find the 
separable state $\sigma$ closest to $\rho$ with
respect to a certain metric,
necessary to define $D_x(\rho)\equiv D_x(\rho, \sigma)$. 
There are several distances to choose from, for instance

\smallskip 
\noindent
{\bf G1. Bures distance} \cite{VP98} 
$D_{\rm B}(\rho)\equiv \min_{\sigma \in {\cal M}_S} D_{\rm B}(\rho,\sigma)$,
\smallskip 

\noindent
{\bf G2. Trace distance} \cite{EAP03} $D_{\rm Tr}(\rho) \equiv \min_{\sigma \in
{\cal M}_S} 
D_{\rm Tr}(\rho,\sigma)$,
\smallskip 

\noindent
{\bf G3. Hilbert-Schmidt distance} \cite{WT99} 
$D_{\rm HS}(\rho) \equiv \min_{\sigma \in {\cal M}_S} 
D_{\rm HS}(\rho,\sigma)$.
\smallskip
 
\noindent
The Bures and the trace metrics are monotone 
(see section 13.2 and 14.1),
which directly implies {\bf (E2a)}, 
while $D_{\rm B}$  fulfils also the stronger property {\bf (E2)} \cite{VP98}.  
Since the HS metric is not monotone \cite{Oz01} it is not at all clear,
whether the minimal Hilbert--Schmidt distance 
is an entanglement monotone \cite{VDM02}.
Since the diameter of the set of mixed states 
with respect to the above distances is finite,
all distance measures cannot satisfy even 
the partial additivity {\bf (E3a)}.

Although quantum relative entropy is not exactly a distance, but rather a 
contrast function, it may also be used to characterise entanglement.
\smallskip

\noindent
{\bf G4. Relative entropy of entanglement} \cite{VPRK97} 
$D_{R}(\rho)\equiv \min_{\sigma \in {\cal M}_S} S(\rho||\sigma)$.

\smallskip
\index{entropy!relative}
\noindent In view of the discussion in chapter 13
this measure has an appealing interpretation as distinguishability 
of $\rho$ from the closest separable state.
For pure states it coincides with the 
 entanglement entropy, $D_{R}(|\phi\rangle)=E_1(|\phi\rangle)$ 
\cite{VP98}. Analytical formulae for $D_{R}$ are known in 
certain cases only \cite{VPRK97,VW01,Is03},
but it may be efficiently computed numerically \cite{RH03}. 
This measure of entanglement is convex (due to double
convexity of relative entropy)  and continuous \cite{DH99},
but not additive \cite{VW01}.
It is thus useful to study the regularised quantity,
$\lim_{n \to \infty} D_{R}(\rho^{\otimes n})/n$.
This limit exists due to subadditivity
of relative entropy and
has been computed in some cases \cite{AEJPVM01,AMVW02}.

\smallskip
\noindent
{\bf G5. Reversed relative entropy of entanglement} 
 $D_{RR}(\rho)\equiv \min_{\sigma \in {\cal M}_S} S(\sigma||\rho)$.

\smallskip
This quantity with exchanged arguments is not so interesting per se,
but its modification $D_{RR}'$ -- the minimal entropy with respect to the 
set ${\cal M}_{\rho}$ of separable states $\rho'$ locally identical to $\rho$, 
$\{ \rho' \in {\cal M}_{\rho}: \rho'_A=\rho_A$ and $\rho'_B=\rho_B \}$,
 provides a distinctive example of 
an entanglement measure, 
 which satisfies  the additivity condition {\bf (E3)} \cite{EAP03}.
(A similar
measure based on modified relative entropy was introduced by Partovi 
\cite{Pa04}.)

\index{robustness} 
\noindent
{\bf G6. Robustness} \cite{VT99}.  $R(\rho)$ measures
the endurance of entanglement by quantifying 
the minimal amount of mixing with separable states needed to wipe
out the entanglement, 
\begin{equation}
R(\rho)\equiv
\min_{\rho_s^- \in {\cal M}_S} \Bigl(\min_{s\ge 0}\ s: \rho_s^+ =\frac{1}{1+s}
                      (\rho+ s \rho_s^-) \in {\cal M}_S \Bigr) \ .
\label{robust}
\end{equation}
As shown if Fig. \ref{fig:ent06}c
the robustness $R$ may be interpreted as
a minimal ratio of the HS distance $1-a=s/(1+s)$ of $\rho$ to the set ${\cal
M}_S$ 
of separable states to the width $a=1/(1+s)$ of this set.
This construction does not depend on the boundary of the
entire set ${\cal M}$, in contrast with the 
best separable approximation.
(In the two-qubit case
the entangled state used for 
BSA (\ref{LSdecomp}) is pure, $\rho_b=|\phi_b\rangle \langle \phi_b|$,
the weight $\Lambda$ is a monotone \cite{EB01},
so the quantity $(1-\Lambda) E_1(\phi_b)$
works as a measure of entanglement \cite{LS98,WK01}).
 Robustness is known to 
be convex and monotone, but is not additive \cite{VT99}.
Robustness for two--qubit states diagonal in the Bell basis
was found in \cite{AJ03}, while a
generalisation of this quantity was proposed in \cite{St03}.

\smallskip
\index{fidelity!maximal} 
\noindent
{\bf G7. Maximal fidelity} $F_{m}$ with respect to the set ${\cal M}_{\rm max}$ 
of maximally entangled states \cite{BVSW96},
 $F_m(\rho)\equiv \max_{\phi \in {\cal M}_{\rm max}} F(\rho,|\phi\rangle\langle
\phi|)$.

Strictly speaking the maximal fidelity cannot be considered as 
a measure of entanglement,
since it does not satisfy even weak discriminance {\bf (E1a)}.
However, it provides a convenient
way to characterize, to what extent $\rho$ may approximate 
a maximally entangled state required for various tasks of quantum information
processing,
so in the two-qubit case it is called the {\it maximal singlet fraction}. 
Invoking (9.31) we see that $F_{m}$ is a function
of the minimal Bures distance from $\rho$
to the set ${\cal M}_{\rm max}$. 
An explicit formula for the maximal fidelity for a two-qubit state
was derived in \cite{BHHH00}, 
while relations to other
entanglement measures were analysed in \cite{VV02b}.

\smallskip
{\centerline {\bf  II. Extensions of pure--state measures}}

Another class of mixed--states entanglement measures can be derived
from quantities characterizing entanglement of pure states.
There exist at least two different ways of proceeding. 
The {\it convex roof} construction 
\cite{Uh98,Uh03}
defines $E(\rho)$ 
as the minimal average quantity
$\langle E(\phi )\rangle$
taken on pure states forming $\rho$.
The most important measure is induced by the entanglement entropy
(\ref{entanentr}).

\index{entanglement!of formation} 
\noindent
{\bf P1. Entanglement of Formation} (EoF) \cite{BVSW96}  
\begin{equation}
 E_F(\rho)\  \equiv \ 
\min_{{\cal E}_{\rho}}  \sum_{i=1}^M p_i E_1(|\phi_i\rangle) \ ,
\label{eof}
\end{equation}
where the minimisation

 is performed over an ensemble 
of all possible decompositions
\begin{equation}
{\cal E}_{\rho}=\{p_i,|\phi_i\rangle\}_{i=1}^M \ : 
\ \rho=\sum_{i=1}^M p_i |\phi_i\rangle \langle \phi_i| 
{\rm \quad with \quad}
p_i>0,  \quad\sum_{i=1}^M p_i =1\ .
\label{ensembl}
\end{equation}
(A dual quantity defined 
by maximisation over ${\cal E}_{\rho}$  
is called {\it entanglement of assistance} \cite{DFMSTU99},
and both of them are 
related to relative entropy of entanglement of
an extended system \cite{HV00}.)

The ensemble ${\cal E}$ for which 
the minimum (\ref{eof}) is realised is called {\it optimal}.
Several optimal ensembles might exist, and  
the minimal ensemble length  $M$ is called the {\it cardinality} of 
the state $\rho$. If the state is separable
then $E_F(\rho)=0$, and the cardinality coincides with the
minimal length of the decomposition  (\ref{mixedsep}).
Due to Carath\'eodory's theorem
the cardinality of $\rho \in {\cal M}^{NK}$
does not exceed the squared rank of the state,
$r^2\le N^2K^2$  \cite{Uh98}.
In the two-qubit case it is sufficient to take
 $M=4$ \cite{Wo98},
and this length is necessary for some states of
rank $r=3$ \cite{AVM01}. In higher dimensions
there exists states for which $M >NK\ge r$ \cite{DTT00}.

Entanglement of formation enjoys several
appealing properties: it may be interpreted as 
the minimal pure--states entanglement 
required to build up the mixed state. 
It satisfies by construction the discriminance 
property {\bf (E1)} and is convex
and monotone \cite{BVSW96}.
EoF is known to be continuous \cite{Ni00b},
and for pure states it is by construction equal to 
the entanglement entropy $E_1(|\phi\rangle)$.
To be consistent with normalisation
{\bf (E6)} one often uses a
rescaled quantity, $E_F' \equiv E_F/\ln 2$.
 
Two other properties are still to be desired, 
if EoF is to be an ideal entanglement measure:
we do not know,  whether EoF is additive
and EoF is not easy to evaluate.
Additivity of
 EoF has been demonstrated in special cases only,
if one of the states is a product state \cite{BN01}, 
is separable  \cite{VW01} or if it is supported on a specific 
subspace \cite{VDC02}. At least
we can be sure that EoF satisfies subadditivity {\bf  (E5b)},
since the tensor product of the optimal
decompositions of $\rho$ and $\sigma$ provides 
an upper bound for $E(\rho \otimes \sigma)$.

Explicit analytical formulae were derived for the two-qubit
 system \cite{Wo98}, and a certain class of symmetric 
states in higher dimensions \cite{TVo00,VW01},
while for the $2 \times K$ systems at least lower bounds are known
 \cite{CLLH02,LBZW03,Ge03}.
Numerically EoF may by estimated by minimisation over the  
space of unitary matrices $U(M)$. A search for
the optimal ensemble can be based on simulated annealing \cite{Zy99}, 
on a faster conjugate--gradient method \cite{AVM01},
or on minimising the conditional mutual information \cite{Tu01}.

\smallskip
\noindent
{\bf P2. Generalised Entanglement of Formation} (GEoF)   
\begin{equation}
 E_q(\rho)\  \equiv \ 
\min_{{\cal E}_{\rho}}  \sum_{i=1}^M p_i \, E_q(|\phi_i\rangle) \ ,
\label{geof}
\end{equation}
where $E_q(|\phi\rangle)=S_q[{\rm Tr}_B(|\phi\rangle \langle \phi|)]$ 
stands for the R{\'e}nyi entropy of entanglement.
 Note that an optimal ensemble for a certain value of $q$ needs 
not to provide the minimum for $q' \ne q$.
GEoF  is asymptotically continuous only in the limit $q \to 1$
for which it coincides with EoF.
In the very same way, the convex roof construction
can be applied to extend any pure states entanglement measure for mixed states.
In fact, several measures introduced so far are related to GEoF. 
For instance, the convex roof extended negativity \cite{LCOK03}
and {\it concurrence of formation} \cite{Woo01,RC03,MKB04}
are related to $E_{1/2}$ and $E_2$, respectively.
\index{concurrence}

There is another way to make use of pure
state entanglement measures.
In analogy to the fidelity between two mixed states, equal to 
the maximal overlap between their purifications,
one may also purify $\rho$ by 
a pure state $|\psi\rangle \in ( {\cal H}_N \otimes {\cal H}_K)^{\otimes 2}$.
Based on the entropy of entanglement (\ref{entanentr}) 
one defines 

\smallskip
\noindent
{\bf P3. Entanglement of purification} \cite{THLV02,BB02}  
\begin{equation}
 E_P(\rho)\ \ \ \equiv \ \ \ \ 
\min_{|\phi\rangle:\ \rho={\rm Tr}_{KN}(|\phi\rangle \langle \phi |)}
  E_1(|\phi\rangle) \ ,
\label{epurif}
\end{equation}
but any other measure of pure states entanglement
might be used instead.

The entanglement of purification is continuous and
monotone under strictly local operations (not under LOCC).
It is not convex, but more importantly,
it does not satisfy even
the weak discriminance {\bf (E1a)}.
In fact $E_P$ measures {\sl correlations}
between both subsystems,
and is positive for any non-product, separable mixed state \cite{BB02}.
Hence entanglement of purification is {\sl not} 
an entanglement measure,  
but it happens to be helpful 
to estimate a variant of the entanglement cost \cite{THLV02}.
To obtain a reasonable measure one needs to allow
for an arbitrary extension of the system size, as assumed by defining 

 \smallskip
\noindent
{\bf P4. Squashed entanglement}
\cite{CW04}  
\begin{equation}
 E_{S} (\rho^{AB})\ \ \ \equiv \ \ \ \ 
\inf_{\rho^{ABE}}
{\frac{1}{2}\bigl[ S(\rho^{AE})+S(\rho^{BE}) - S(\rho^E}) -S(\rho^{ABE}) \bigr]
\ ,
\label{esquashed}
\end{equation}
where the infimum is taken over all extensions $\rho^{ABE}$
of an unbounded size such that
${\rm Tr}_{E}(\rho^{ABE})=\rho^{AB}$.
Here $\rho^{AE}$ stands for ${\rm Tr}_{B}(\rho^{ABE})$ while 
$\rho^{E} ={\rm Tr}_{AB}(\rho^{ABE})$. 
Minimized quantity is proportional
to quantum conditional mutual information of $\rho^{ABE}$ \cite{Tu01}
and its name refers to 'squashing out'  the classical correlations.

Squashed entanglement
is convex, monotone and  vanishes
for every separable state \cite{CW04}. 
If $\rho^{AB}$ is pure then $\rho^{ABE}=\rho^E\otimes \rho^{AB}$,
hence $E_{S} =[S(\rho^{A})+S(\rho^{B})]/2 =S(\rho^{A})$
and the squashed entanglement reduces to  
the entropy of entanglement $E_1$.
It is charactarised by asymptotic
continuity \cite{AF04}, and additivity {\bf (E5)}, which  
is a consequence of the strong subadditivity of the von Neumann entropy
\cite{CW04}.
Thus $E_S$ would be a perfect measure of entanglement, if we only knew
how to compute it...
 
\smallskip
{\centerline {\bf  III. Operational measures}}

Entanglement may also be quantified in an abstract manner
by considering the minimal resources required to generate a given state
or the maximal entanglement yield. 
These measures are defined implicitly, since one
deals with an infinite set of copies of the state analysed 
and assumes an optimisation over all possible LOCC protocols.

\smallskip
\noindent
{\bf O1.  Entanglement cost} \cite{BVSW96,Ra99} $E_C(\rho)=\lim_{n\to
\infty}\frac{m}{n}$
where $m$ is the number of singlets $|\psi^-\rangle$ needed to produce
locally $n$ copies of the analysed state $\rho$.

Entanglement cost has been calculated for instance 
for states supported on a subspace such that tracing out one of the parties
forms an entanglement breaking channel (super  separable map)  \cite{VDC02}.
Moreover, entanglement cost was shown \cite{HHT01} to be  equal to the  
regularised  entanglement of formation,  
$E_C(\rho)=\lim_{n\to \infty} E_F(\rho^{\otimes n})/n$.
Thus, if we knew that EoF is additive,
the notions of entanglement cost and entanglement of formation would
coincide...

\noindent
{\bf O2. Distillable entanglement} \cite{BVSW96,Ra99}. $E_D(\rho)= 
\lim_{n\to \infty}\frac{m}{n}$
where $m$ is the maximal number of singlets $|\psi^-\rangle$ obtained 
out of $n$ copies of the state $\rho$ by an optimal LOCC conversion protocol.

Distillable entanglement is a measure of a fundamental importance,
since it tells us how much entanglement one may extract out of 
the state analysed and use e.g. for the cryptographic purposes.
It is rather difficult to 
compute, but there exist analytical bounds due to Rains 
\cite{Ra99b,Ra01},
and an explicit optimisation formula was found \cite{DW05}.
$E_D$ is not likely to be  convex \cite{SST01},
although it satisfies the weaker condition {\bf (E3a)} \cite{DHR02}.

\medskip
{\centerline {\bf  IV. Algebraic measures}}

\smallskip
\noindent 
If a partial transpose of a state $\rho$ is not positive then 
$\rho$ is entangled due to the PPT criterion {\bf B1}.
The partial transpose preserves the trace, so if $\rho^{T_A}\ge 0$
then $||\rho^{T_A}||_{\rm Tr}={\rm Tr}\rho^{T_A}=1$.
Hence we can use the trace norm to characterize the degree, 
to which the positivity of $\rho^{T_A}$ is violated.

\smallskip
\noindent
{\bf N1. Negativity} \cite{ZHSL98,EP99}, \  
 ${\cal N}_T(\rho)\equiv   ||\rho^{T_A}||_{\rm Tr} - 1$.
\smallskip

\index{negativity}
Negativity is easy to compute,
convex (partial transpose is linear and the trace norm is convex)
and monotone \cite{Ei01,ViWe02}.
It is not additive, but this drawback may be  cured by 
defining the {\it log--negativity}
$\ln  ||\rho^{T_A}||_{\rm Tr}$. 
It was used by Rains \cite{Ra01} to obtain bounds
on distillable entanglement.
However, the major deficiency of the negativity
is its failure to satisfy {\bf (E1)}  -- by construction 
${\cal N}_T(\rho)$ cannot detect PPT bound entangled states.
In the two--qubit case the spectrum of $\rho^{T_A}$
contains at most a single negative eigenvalue  
\cite{STV98},
so ${\cal N}'_T\equiv {\rm max} \{0,-2\lambda_{\rm min}\}={\cal N}_T$.
This observation  explains the name on the one hand,
and on the other provides a geometric interpretation:
${\cal N}_T'(\rho)$ measures the minimal relative weight
of the maximally mixed state $\rho_*$ which needs to be 
mixed with $\rho$ to produce a separable mixture \cite{VDM02}.
In higher dimensions several eigenvalues of the partially transposed state may
be negative,
so in general  ${\cal N}_T \ne {\cal N}'_T$, and the latter quantity
is proportional to complete co--positivity
(11.13) of a map $\Phi$ associated with the state $\rho$.

Negativity is not the only application of the trace norm \cite{ViWe02}.
Building on the positive maps criterion {\bf A1} for separability 
one might analyse analogous quantities
for any (not completely) positive map,
${\cal N}_{\Phi}(\rho)\equiv   ||(\Phi \otimes {\mathbbm 1}) \rho||_{\rm Tr} -
1$.
Furthermore, one may consider another quantity related to the 
reshuffling criterion {\bf B6}.

\noindent
{\bf N2. Reshuffling negativity}  \cite{CW02,Rud03}.
 ${\cal N}_R(\rho)\equiv   ||\rho^R||_{\rm Tr} - 1$.

This quantity is convex due to linearity of reshuffling
and non-increasing under local measurements,
but may increase under partial trace \cite{Rud03}.
For certain bound entangled states ${\cal N}_R$ is positive; 
unfortunately not for all of them...
A similar quantity with the minimal cross-norm
$||\cdot ||_{\gamma}$ was studied by Rudolph \cite{Rud01},
who showed that  ${\cal N}_{\gamma}(\rho)=||\rho||_{\gamma} - 1$
is convex and monotone under local operations.
However, 
 $||\rho^R||_{\rm Tr}$ is easily computable
from the definition (10.33), in contrast to $||\rho||_{\gamma}$.

{\small
\begin{table}
\index{entanglement!measures} 
\index{maps!unital} 
\index{maps!bistochastic} 
\caption{Properties of entanglement measures: 
discriminance E1,
mono\-to\-nicity E2, 
convexity E3,
asymptotic continuity E4,
additivity E5, extensivity E5a,
computability E7: explicit closed formula C, optimisation
over a finite space F or an infinite space I; 
R{\'e}nyi parameter $q$ for pure states.
}
 \smallskip
\hskip -0.5cm
{\renewcommand{\arraystretch}{1.43}
\begin{tabular}
[c]{lcccccccc} 
\hline \hline
computab   q 
Entanglement measure  &
               {\bf E1} &  {\bf E2} & {\bf E3} &  {\bf E4} &  {\bf E5} & {\bf
E5a} &  {\bf E7} &  $q$ \\
  \hline \hline
{\bf G1} Bures distance  $D_{B}$ &
          Y &   Y  & Y   &   N   &  N  &   N    &  F   & $2$ \\
{\bf G2} Trace distance  $D_{\rm Tr}$ &
          Y &   Y  & Y   &   N  &  N  &   N    &  F   &   \\
{\bf G3} HS distance     $D_{\rm HS}$ &
          Y &   ?  & ?   &   N  &  N  &   N    &  F   &  \\
 {\bf G4} Relative entropy $D_{R}$ &
          Y &   Y  & Y   &   Y  &   N    &  ?   & F   & $1$ \\
 {\bf G5'} reversed RE, $D_{RR}'$ &
          ? &   Y  & Y   &   Y   &  Y  &   Y    &  F   & $1$ \\
 {\bf G6} Robustness $R$ &
          Y &   Y  & Y   &   N   &  N &   N    &  F   & $1/2$ \\
\hline
{\bf P1}  Entangl.  of formation $E_F$ &
          Y &   Y &   Y  &   Y    &  ?   & ? &   C/F   &  $1$ \\
{\bf P2}  Generalised EoF, $E_q$ &
          Y &   Y &   ?  &   N    &  ?   & ? &   F  &  $q$ \\
{\bf P4}  Squashed entangl.  $E_S$ &
           ? &   Y &   Y  &   Y    &  Y   & Y &  I  & $ 1$ \\
\hline
{\bf O1} Entan. cost $E_C$ &
          Y &   Y &   Y  &    Y    &  ?   & Y &   I  &  $1$ \\
{\bf O2}  Distillable  entangl. $E_D$ &
          N &   Y &   N(?) &   Y  &   ?  & Y  &    I   &  $1$ \\
\hline
{\bf N1}  Negativity   ${\cal N}_T$ &
          N  &  Y &    Y  &   N    &  N   & N &   C  &  $1/2$ \\
{\bf N2}   Reshuffling negativity   ${\cal N}_R$ &
          N &   N &    Y  &   N    &  N   & N &    C   &  $1/2$ \\
\hline \hline
\end{tabular}
}
\label{tab:entmeas}
\end{table}
} 

\medskip 

There exists 
also attempts  to quantify entanglement
by the dynamical properties of a state and the 
speed of decoherence \cite{BJO01},
or the secure key distillation rate \cite{HHHO05,DW05}
 and several others...
For multipartite systems the problem 
gets even more demanding \cite{CKW00,EB01,WG03,Pa04}.

We end this short tour through the vast garden
of entanglement measures, 
by studying how they behave for pure states.
Entanglement of formation and purification 
coincide by construction with the entanglement entropy $E_1$.
So is the case for both operational measures,
since conversion of $n$ copies of an analyzed pure state into
$m$ maximally entangled states is reversible.  
The negativities are easy to compute. 
For any pure state $|\psi\rangle\in {\cal H}_N \otimes {\cal H}_N$
 written in the Schmidt form,  
the non-zero entries of the density matrix $\rho_{\psi}$ of size $N^2$
are equal to $\sqrt{\lambda_i\lambda_j}$, $i,j=1,\dots,N$.  
The reshuffled matrix $\rho_{\psi}^{R}$ becomes diagonal,
while partially transposed matrix $\rho_{\psi}^{T_2}$ has a block structure:
it consists of  $N$ Schmidt  components $\lambda_i$ at the diagonal
which sum to unity 
 and $N(N-1)/2$ off-diagonal blocks of size $2$,
one for each pair of different indices $(i,j)$.
Eigenvalues of each block are
 $\pm \sqrt{\lambda_i \lambda_j}$,
so both traces norms are equal to the sum of all entries.
Hence both negativities coincide for pure states,
\begin{equation} 
{\cal N}_T(\rho_{\psi})=
{\cal N}_R(\rho_{\psi})=
\sum_{i,j=1}^N \! \!  \sqrt{\lambda_i \lambda_j } -1 =
 \bigr( \sum_{i}^N \! \! \sqrt{\lambda_i}\bigl)^2  
 - 1 =  e^{E_{1/2}}-1 \ , 
\label{negatpur}
\end{equation}
and vary from $0$ for separable states to $N-1$ 
for maximally entangled states.
Also maximal fidelity and robustness for pure states
become related,  $F=\exp(E_{1/2})/N$ 
and 
$R= {\cal N}_T=\exp(E_{1/2})-1$ \cite{VT99}.
On the other hand, the minimal Bures distance to the closest
separable (mixed) state \cite{VP98}
becomes a function of the R{\'e}nyi entropy of order two,
\begin{equation} 
D_{\rm B}(\rho_{\phi})= \bigl( 2-2\sum_{i=1} \lambda_i^2\bigr)^{1/2}
   = \sqrt{ 2-2 e^{-E_2/2} }\ , 
\label{Burpure}
\end{equation}
and is equal to concurrence (\ref{tangle}), 
while the Bures distance to the closest separable pure state
$D_{\rm B}^{\rm pure}(|\phi\rangle)= [2(1-\sqrt{1-\lambda_{max}})]^{1/2}$
is a function of  $E_{\infty}=-\ln \lambda_{\rm max}$.
\index{entropy!R{\'e}nyi}
The  R{\'e}nyi parameters $q$ characterising behaviour 
of the discussed  measures of entanglement for pure states 
are collected in Table \ref{tab:entmeas}.

Knowing that a given state $\rho$ can be locally 
transformed into $\rho'$ implies that $E(\rho)\ge E(\rho')$
for any measure $E$, but the converse is not true.
Two R{\'e}nyi entropies of entanglement of 
different (positive) orders generate different order 
in the space of pure states. 
By continuity this is also the case
for mixed states, and the relation 
\begin{equation} 
E_A(\rho_1) \: \le \:  E_A(\rho_2) 
 \ \ \Leftrightarrow \ \
E_B(\rho_1) \: \le \: E_B(\rho_2) 
\label{entorder}
\end{equation}
does not hold.
For a certain pair of mixed states 
it is thus likely that one state is more entangled
with respect to a given measure, while the other one,
with respect to another measure of entanglement \cite{EP99}. 
If two measures $E_A$ and $E_B$ coincide for pure states
they are identical or they {\sl do not}
generate the same order in the set of mixed states \cite{VP00}.
Hence entanglement of formation and distillable
entanglement do not induce the same order.
On the other hand,  
several entanglement measures are correlated
and knowing $E_A$ one may try to find lower and upper bounds for $E_B$.

The set of entanglement measures shrinks,
if one imposes even some of the desired properties {\bf (E1)-(E7)}.
The asymptotic continuity {\bf (E4)}
is particularly restrictive.
For instance, among generalized R{\'e}nyi entropies it is 
satisfied only by 
the entropy of entanglement $E_1$ \cite{Vi00}.
If a measure $E$ satisfies additionally
monotonicity {\bf (E2a)} under deterministic LOCC 
and extensivity {\bf (E5a)}, it  is bounded by 
the distillable entanglement and entanglement cost \cite{HHH00,HoM01},
\begin{equation} 
E_D(\rho) \ \le \ E (\rho) \  \le \  E_C(\rho) \ . 
\label{entbounds}
\end{equation}
Interestingly, the two first measures introduced in the pioneering
paper by Bennett et al. \cite{BVSW96} occurred to be
the extreme entanglement measures. For pure states both 
of them coincide, and we arrive at a kind of {\bf uniqueness theorem}:
{\sl Any  monotone, extensive and  asymptotically continuous
entanglement measure coincides for pure states with the  
entropy of formation} $E_F$ \cite{PR97,Vi00,DHR02}.
This conclusion concerning pure states entanglement 
of bipartite systems may also be reached by an abstract,
thermodynamic approach \cite{VK02}.
 
Let us try to recapitulate the similarities and differences
between four classes of entanglement measures.
For a geometric measure or an extension of a pure states measure
it is not simple to check, which of the desired properties are satisfied.
Furthermore, to evaluate it for a typical mixed state  
one needs to perform a cumbersome optimisation scheme.
One should not expect the remarkable analytical result 
of Wootters \cite{Wo98} for entanglement of formation
in the $2 \times 2$ system,
to be extended for the general $N \times N$ problem,
since  even stating the separability is known to be 
an algorithmically complex task \cite{Gu03}.

Operational measures are nice, especially from the  point of view
of information science,
and extensivity and monotonicity  are 
direct consequence of their definitions. However, they are extremely hard to 
compute. In contrast, algebraic measures are easy to calculate,
but they fail to detect entanglement for all
non--separable states. Summarising, several different measures
of entanglement are thus likely to be still used in future...

\section{Two qubit mixed states}
\label{sec:twomix}

Before discussing the entanglement of two--qubit mixed states
let us recapitulate, in what sense the case $N=2$ differs from $N\ge 3$.

\smallskip
{\bf A) \ algebraic properties}
 
\noindent
{\bf i)}  \ \ $SU(N)\times SU(N)$ is homomorphic to $ SO(N^2)$ for $N=2$ only, 

\noindent
{\bf ii)} \ $SU(N)  \cong    SO(N^2-1)$ for $N=2$ only,
 
\noindent
{\bf iii)} All positive maps  $\Phi:{\cal M}^{(N)} \to {\cal M}^{(N)}$ are
decomposable
for $N=2$ only,

\smallskip
{\bf B) \ $N$--level mono-partite systems} 

\noindent
{\bf iv)} Boundary  $\partial {\cal M}^{(2)}$ consists of pure states only,
 
\noindent \hangafter=1 \hangindent=0.8cm
{\bf v)} \ For any state $\rho_{\vec \tau}$ $\in {\cal M}^{(N)}$ also 
 its antipode 
 $\rho_{-{\vec \tau}}=2\rho_* - \rho_{\vec \tau}$ 
forms a state and there exists a universal NOT operation for $N=2$ only.

\noindent \hangafter=1 \hangindent=0.9cm
{\bf vi)} \ ${\cal M}^{(N)}\subset {\mathbb R}^{N^2-1}$ forms a 
    ball for $N=2$ only,

\smallskip
{\bf C) \ $N\times N $ composite systems} 

\noindent \hangafter=1 \hangindent=0.8cm
{\bf vii)}  For any pure state $|\psi\rangle \in {\cal H}_N \otimes {\cal H}_N$
     there exist $N-1$ independent Schmidt coefficients $\lambda_i$. For $N=2$
     there exists only one independent coefficient $\lambda_1$, hence 
     all entanglement measures are equivalent.

\noindent \hangafter=1 \hangindent=0.9cm
{\bf viii)} The maximally entangled states form the manifold $SU(N)/{\mathbb
Z}_N$,
    which is equivalent to the real projective space ${\mathbb R}{\bf
P}^{N^2-1}$ 
     only for $N=2$.

\noindent
{\bf ix)} \ \ All PPT states of a $N \times N$ system are separable for $N=2$
only. 

\noindent \hangafter=1 \hangindent=0.9cm
{\bf x)}  \ \ \ For any two--qubit mixed state its optimal decomposition
consists
           of pure states of equal concurrence. Thus entanglement of formation 
            becomes a function
            of concurrence of formation for $2 \times 2$ systems.
\smallskip

These features 
demonstrate why 
entanglement of  two-qubit systems is special \cite{VW00}.
Several of these issues are closely related. We have already learned that
decomposability {\bf iii)} is a consequence of {\bf ii)} and implies the
separability {\bf ix)}.
We shall see now, how a group-theoretic fact {\bf i)} allows one to 
to derive a closed formula for EoF of a two qubit system.
We are going to follow the seminal paper of Wootters \cite{Wo98},
who built up on his earlier work with Bennett et al. \cite{BVSW96}
and the paper with Hill  \cite{HW97}.

Consider first a two--qubit pure state $|\psi\rangle$.
Due to its normalisation
the Schmidt components---eigenvalues of the
matrix $\Gamma \Gamma^{\dagger}$ ---satisfy $\mu_1+\mu_2=1$.
The  tangle of  $|\psi\rangle$,
 defined in (\ref{tangle}),  reads
\begin{equation} 
\tau = C^2 =2(1-\mu_1^2 -\mu_2^2) = 4 \mu_1 (1-\mu_1)= 4 \mu_1\mu_2 \ ,
\label{conc1}
\end{equation}
and implies that concurrence is proportional
to the determinant of (\ref{SchmidtC2}),
\begin{equation} 
C = 2 \ |\sqrt{\mu_1\mu_2}|= 2 \ | {\rm det} \Gamma| \ .
\label{conc2}
\end{equation}
\index{concurrence}
Inverting this relation we find the
 entropy of entanglement $E$ as a function of concurrence
\begin{equation} 
E=S(\mu_1,1-\mu_1)
{\quad \rm where \quad}
\mu_1 = \frac{1}{2} \bigl( 1 - \sqrt{1-C^2} \bigr)   
\label{conc3}
\end{equation}
and $S$ stands for the Shannon entropy function, $-\sum_i \mu_i \ln \mu_i$.

Let us represent $|\psi\rangle$ in a particular basis consisting of four
Bell states 
\begin{eqnarray} 
|\psi\rangle & =&  \Bigl[ a_1 |\phi^+\rangle + a_2 i|\phi^-\rangle
 + a_3 i|\psi^+\rangle + a_4 |\psi^-\rangle \Bigr] \\
 & = & \frac{1}{\sqrt{2}}
\Bigl[ (a_1+ia_2) |00\rangle +
 (ia_3+a_4) |01\rangle +
(ia_3-a_4) |10\rangle +
(a_1-ia_2) |11\rangle  \Bigr] \ . \nonumber
\label{magic1}
\end{eqnarray}
Calculating the determinant in Eq. (\ref{conc2})
we find that
\begin{equation} 
C(|\psi\rangle) \: = \: \bigl| \sum_{k=1}^4 a_k^2 \bigr| \ .  
\label{conc4}
\end{equation}
If all its coefficients of $|\psi\rangle$ 
in the basis (\ref{magic1}) are real
then $C(|\phi\rangle)=1$ and the state is maximally entangled. 
This property holds also if we act on 
$|\psi\rangle$ with an orthogonal gate $O \in SO(4)$
and justifies referring to 
(\ref{magic1}) as the {\it magic basis} \cite{BVSW96}.
Any two qubit unitary gate, which is represented 
in it by a real $SO(4)$ matrix  corresponds to a 
local operation
and its action does not influence  entanglement. 
This is how the property {\bf i)} enters the game.
A gate represented by
an orthogonal matrix with det$O=-1$, corresponds to 
SWAP of both qubits. It does not influence the entanglement but is 
non--local \cite{VW00}.

To appreciate another feature of the magic basis
consider the transformation
$|\psi\rangle \to |{\tilde\psi}\rangle=(\sigma_y \otimes
\sigma_y)|\psi^*\rangle$,
in which complex conjugation is taken in the standard basis,
$\{ |00\rangle,|01\rangle,|10\rangle,|11\rangle \}$.
It represents flipping of both spins of the system.
If a state is expressed in the magic basis
this transformation is realized just by complex conjugation.
Expression (\ref{conc4}) implies then 
\begin{equation} 
C(|\phi\rangle) \  = \ | \langle \psi |{\tilde\psi}\rangle |.
\label{conc5}
\end{equation}

\noindent
Spin flipping of mixed states 
is also realised by complex conjugation, 
if $\rho$ is expressed in magic basis. 
Working in standard  basis this transformation reads
\begin{equation} 
\rho \ \to \ {\tilde \rho} \ = \ (\sigma_y \otimes \sigma_y) \rho^* 
 (\sigma_y \otimes \sigma_y)  \ .  
\label{spinflip}
\end{equation}
In the Fano form (\ref{Fano})
flipping corresponds to reversing the signs
of both Bloch vectors,
\begin{equation} 
{\tilde \rho} =  \frac{1}{4} \Bigl[
{\mathbbm 1}_{4} -
\sum_{i=1}^{3}\tau^{A}_i \sigma_i \otimes {\mathbbm 1}_2 -
\sum_{j=1}^{3} \tau^{B}_j   {\mathbbm 1}_2 \otimes \sigma_j +
\sum_{i,j=1}^{3} \beta_{ij} \sigma_i \otimes \sigma_j 
\Bigr] \ .
\label{FanoAB}
\end{equation}
\index{Fano form} 
Root fidelity between $\rho$ 
and $\tilde \rho$
is given by the trace of the positive matrix
 \begin{equation} 
\sqrt{F} \ = \ \sqrt{ \sqrt{\rho} {\tilde \rho} \sqrt{\rho}} \ .  
\label{RRR}
\end{equation}
Let us denote by $\lambda_i$ 
the decreasingly ordered eigenvalues of $\sqrt{F}$,
(singular values of $\sqrt{\rho}\sqrt{\tilde \rho}$). 
The {\it concurrence} of a two-qubit mixed state is now defined  by 
 \begin{equation} 
C(\rho) \ \equiv \ \max \{0,\ \lambda_1-\lambda_2-\lambda_3-\lambda_4\} \ .   
\label{cmixed}
\end{equation}
The number of of positive eigenvalues cannot be greater
then rank $r$ of $\rho$. For a pure state the above definition is thus
 consistent with  $C=| \langle \psi |{\tilde\psi}\rangle |$ and 
expression (\ref{conc4}).

Consider a generic mixed state of full rank
given by its eigendecomposition  $\rho=\sum_{i=1}^4 |w_i\rangle \langle w_i|$.
The eigenstates are subnormalised in a sense that
 $||w_i\rangle|^2$ is equal to the $i$-th eigenvalue $d_i$.
The flipped states $|\tilde w_i\rangle$ are also eigenstates of ${\tilde
\rho}$. 
Defining a symmetric matrix $W_{ij}\equiv \langle w_i|{\tilde w}_j\rangle$
we see that the spectra of $\rho {\tilde \rho}$ and $WW^{*}$ coincide.
Let $U$ be a unitary matrix diagonalising the Hermitian matrix $WW^{*}$.

 Other decompositions of $\rho$
may be obtained by the  Schr\"odinger's mixture theorem (9.33).
\index{theorem!Schr\"odinger} 
In particular, the unitary matrix $U$ defined above
gives  a decomposition into four states 
$|x_i\rangle \equiv \sum_i U^*_{ij}| w_j\rangle$.
They fulfill
\begin{equation} 
\langle x_i | {\tilde x_j}\rangle =(U W U^T)_{ij}=\lambda_i \delta_{ij} \ .   
\label{xxtilde}
\end{equation}
Since $W$ is symmetric, an appropriate choice of phases of eigenvectors 
forming $U$ assures that the diagonal elements of  $U W U^T$ 
are equal to the square roots of the eigenvalues of $WW^{*}$,
which coincide with the  eigenvalues $\lambda_i$ of 
$\sqrt{F}$. 

\begin{figure} [htbp]
  \begin{center} \
 \includegraphics[width=10cm,angle=0]{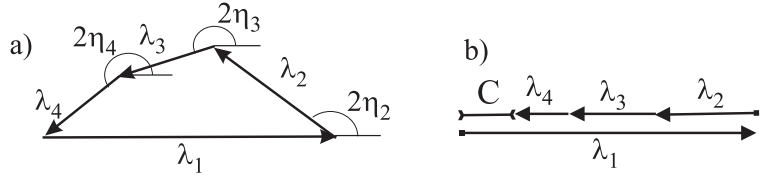}
\caption{Concurrence polygon:
a) quadrangle for a separable state, 
b) line for an entangled state with concurrence $C$.} 
\label{fig:ent07}
\end{center}
 \end{figure}

We are going to show that a state is separable
if $C=0$. Hence $\lambda_1<\lambda_2+\lambda_3+\lambda_4$
and it is possible to find four phases $\eta_i$ such that 
 \begin{equation} 
\sum_{j=1}^4 e^{2\eta_j} \lambda_j  = 0   
\label{etamix}
\end{equation}
In other words such a chain of four
 links of length $\lambda_i$
may be closed to form a polygon
as sketched in Fig.  \ref{fig:ent07}a, in which the phase $\eta_1$
is set to zero. Phases $\eta_i$ allow us to write four
 other pure states
\begin{eqnarray} 
|z_1\rangle = \frac{1}{2}\bigl( 
e^{i\eta_1} |x_1\rangle +e^{i\eta_2} |x_2\rangle +
e^{i\eta_3} |x_3\rangle +e^{i\eta_4} |x_4\rangle \bigr) \ , \nonumber \\
|z_2\rangle = \frac{1}{2}\bigl( 
e^{i\eta_1} |x_1\rangle +e^{i\eta_2} |x_2\rangle -
e^{i\eta_3} |x_3\rangle -e^{i\eta_4} |x_4\rangle \bigr) \ , \\
|z_3\rangle = \frac{1}{2}\bigl( 
e^{i\eta_1} |x_1\rangle -e^{i\eta_2} |x_2\rangle +
e^{i\eta_3} |x_3\rangle -e^{i\eta_4} |x_4\rangle \bigr) \ , \nonumber \\
|z_4\rangle = \frac{1}{2}\bigl( 
e^{i\eta_1} |x_1\rangle -e^{i\eta_2} |x_2\rangle -
e^{i\eta_3} |x_3\rangle +e^{i\eta_4} |x_4\rangle \bigr) \ . \nonumber 
\label{sepdecomp}
\end{eqnarray}
On the one hand they form a decomposition of the state analysed,
 $\rho=\sum_i |x_i\rangle \langle x_i| =\sum_i |z_i\rangle \langle z_i|$.
On the other hand,  due to (\ref{xxtilde},\ref{etamix})
$\langle z_i |{\tilde z}_i\rangle=0$ for $i=1,...,4$, 
hence each pure state $|z_i\rangle$ of the decomposition is separable
and so is $\rho$.

Consider now a mixed state $\rho$ for which $C>0$ 
since $\lambda_1$ is so large that the chain cannot be closed 
-- see Fig.  \ref{fig:ent07}b.
Making use  of the pure states $|x_i\rangle$
constructed before we introduce a set of four states
 \begin{equation} 
|y_1\rangle =|x_1\rangle\ , \quad 
|y_2\rangle =i|x_2\rangle\ , \quad 
|y_3\rangle =i|x_3\rangle\ , \quad 
|y_4\rangle =i|x_4\rangle\ . \quad 
\label{yyyy}
\end{equation}
and a symmetric 
matrix $Y_{ij}=\langle y_i | {\tilde y_j}\rangle$, the 
relative phases of which are chosen in such a way
that relation (\ref{xxtilde}) implies
\begin{equation} 
{\rm Tr}\ Y =\sum_{i=1}^4 
\langle y_i | {\tilde y_j}\rangle 
= \lambda_1-\lambda_2-\lambda_3-\lambda_4 \ = \ C(\rho) \ . 
\label{yyy2}
\end{equation}
The states (\ref{yyyy}) are subnormalised,
hence the above expression represents an average
of a real quantity, the absolute value of which 
coincides with the concurrence (\ref{conc5}).
Using Schr\"odinger's  theorem again 
one may find yet another decomposition, 
$|z_i\rangle \equiv \sum_i V^*_{ij}| y_j\rangle$,
such that every state 
has {\sl the same} concurrence, $C(|z_i\rangle)=C(\rho)$
for $i=1,\dots,4$. 
To do so define a symmetric matrix
$Z_{ij}=\langle z_i | {\tilde z_j}\rangle$
and observe that Tr$Z={\rm Tr} (VYV^T)$. 
This trace does not change if $V$ is real 
and $V^T=V^{-1}$ follows. 
Hence one may find such an orthogonal $V$
that all overlaps are equal to concurrence, $Z_{ii}=C(\rho)$,
and produce the final decomposition 
$\rho=\sum_i |z_i\rangle \langle z_i|$.

The above decomposition is optimal
for the concurrence of formation \cite{Wo98}, 
\begin{equation}
C_F(\rho) \ = \
\min_{{\cal E}_{\rho}}  \sum_{i=1}^M p_i C(|\phi_i\rangle) 
\ = \ \min_{V}  \sum_{i=1}^m |(VYV^T)_{ii}| \ ,
\label{concform}
\end{equation}
where the minimum over ensembles 
${\cal E}_{\rho}$ may be replaced by a minimum
over $m \times n$ rectangular matrices $V$,
containing $n$ orthogonal vectors of size $m \le 4^2$ \cite{Uh98}.
The function 
relating $E_F$ and concurrence is convex,
hence the decomposition into pure states
of equal concurrence is also optimal for entropy of formation.
Thus $E_F$ of any two-qubit state $\rho$
 is given by the function (\ref{conc3}) 
of concurrence of formation $C_F$  equal to $C(\rho)$, and 
defined by (\ref{cmixed}).
A streamlined proof of this fact 
was provided in \cite{AVM01}, while 
the analogous problem for two rebits was solved in \cite{CFR02}.

While the algebraic fact {\bf i)} was used to 
calculate concurrence, 
the existence of a general formula for maximal fidelity 
hinges on property {\bf ii)}.  
Let us write the state $\rho$ in its Fano form 
(\ref{Fano}) and analyze invariants of local
unitary transformations $U_1\otimes U_2$.
Due to the relation $SU(2)\approx SO(3)$
this transformation may be interpreted as 
an independent rotation of both Bloch vectors,
${\vec \tau}^{A} \to O_1 {\vec \tau}^{A}$
and
${\vec \tau}^{B} \to O_2 {\vec \tau}^{B}$.
Hence the real correlation matrix
$\beta_{ij}={\rm Tr}(\rho \sigma_i \otimes \sigma_j)$
may be brought into diagonal form $K=O_1 \beta O_2^T$. 
The diagonal elements may admit negative values 
since we have restricted orthogonal matrices to fulfill
det$O_i=+1$. Hence $|K_{ii}|=\kappa_i$
where $\kappa_i$ stand for singular values  of $\beta$. 
Let us order them decreasingly.
\index{local invariants} 
By construction they are invariant with respect to 
local unitaries,
and govern the maximal fidelity with respect to 
maximally entangled states \cite{BHHH00},
\index{Fano form}
\index{fidelity!maximal} 
\begin{equation} 
F_m(\rho) =\frac{1}{4} \bigl[
1+\kappa_1+\kappa_2 - {\rm Sign}[{\rm det}(\beta)] \ \kappa_3 \bigr] \ .
\label{maxfid2}
\end{equation}
Two--qubit density matrix 
is specified by $15$ parameters, the local unitaries
are characterized by $6$ variables, so there exist
$9$ functionally independent local invariants.
However, two states are locally equivalent
if they share additional $9$ discrete invariants
which determine signs of $\kappa_i$, $\tau^A_i$
and $\tau^B_i$ \cite{Ma02}. 
A classification of mixed states based on degeneracy
and signature of $K$ was worked out in \cite{GRB98,EM01,KZ01}.

It is instructive to compute
explicit formulae for above entanglement measures
 for several families of two--qubit states.
The concurrence of a Werner state (\ref{Wernern})
is equal to the negativity,   
\begin{equation} 
  C\bigl( \rho_W(x)\bigr)={\cal N}_T\bigl( \rho_W(x)\bigr) =\left\{
\begin{array}{ccc}
  0   \quad & {\rm if } \quad \quad & x \le 1/3  \\
 (3x-1)/2       \quad & {\rm if } \quad \quad & x \ge 1/3 \\
\end{array} \right. \ , 
\label{negwern}
\end{equation}
its entanglement of formation is given by (\ref{conc3}),
while $F_m=(3x-1)/4$.

Another interesting family of states arises as a 
convex combination 
of a Bell state with an orthogonal separable state  \cite{HHH00b}
\begin{equation}
\sigma_H(a) \  \equiv \ a \, |\psi^-\rangle \langle \psi^-|  
 \ + \ (1-a)\,  |00\rangle \langle 00| \ .
\label{sigmac}
\end{equation}
Concurrence of such a state is by construction
equal to its parameter, $C=a$,
while the negativity reads  ${\cal N}_T=
\sqrt{(1-a)^2+a^2} +a-1$. The relative  
entanglement of entropy reads \cite{VP98} 
$E_R=(a-2) \ln (1-a/2)+(1-a)\ln (1-a)$. 
We shall use also a mixture of Bell states, 
\begin{equation}
\sigma_B(b)\  \equiv \ b \, |\psi^-\rangle \langle \psi^-|
+(1-b) \,   |\psi^+\rangle \langle \psi^+| \  ,
\label{sigmab}
\end{equation}
for which by construction  $F_m=\max\{b,1-b\}$
and  $C={\cal N}_T= 2F_m-1$. 

Entanglement measures are correlated:
they vanish for separable states and
coincide for maximally entangled states.
For two--qubit systems several explicit bounds are known.
Concurrence forms an upper bound for negativity \cite{EP99,Zy99}.
This statement was proved in \cite{VADM01},
where it was shown that these measures coincide 
if the eigenvector of $\rho^{T_A}$ corresponding to the negative
eigenvalue is maximally entangled. The lower bound
\begin{equation} 
 C \ \ge \ {\cal N}_T \ \ge \ \sqrt{(1-C)^2+C^2} + C-1 
\label{concneg}
\end{equation}
 is achieved  \cite{VADM01} for states (\ref{sigmac}).

%
\begin{figure} [htbp]
  \begin{center} \
 \includegraphics[width=12.8cm,angle=0]{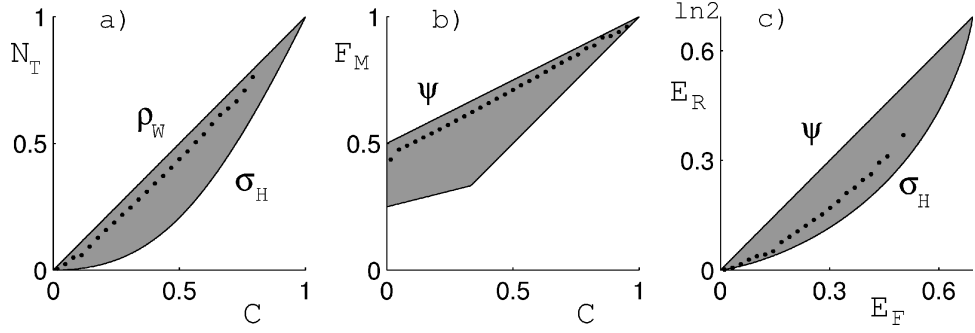}
\caption{Bounds between entanglement measures for 
two--qubits:
a) negativity versus concurrence (\ref{concneg}), 
b) maximal fidelity versus concurrence (\ref{concfid}),
c) relative entropy of entanglement
 versus entanglement of formation  (\ref{eforerel}).
Labels represent families of 
extremal states while 
dots denote averages taken with respect to the HS measure
in ${\cal M}^{(4)}$.} 
\label{fig:ent08}
\end{center}
 \end{figure}

\index{negativity}
\index{concurrence}
Analogous tight bounds between maximal fidelity 
and concurrence or negativity were established in 
\cite{VV02b}, 
\begin{equation} 
 \frac{1+C}{2}  \ge F_m \ge 
\left\{
\begin{array}{ccc}
 (1+C)/4  \quad & {\rm if } \quad \quad & C \le 1/3  \\
  C         \quad & {\rm if } \quad \quad & C \ge 1/3 \\
\end{array} \right. \ , 
\label{concfid}
\end{equation}
\begin{equation} 
 \frac{1+{\cal N}_T}{2}  \ge F_m \ge 
\left\{
\begin{array}{ccc}
 \frac{1}{4}+\frac{1}{8}
\Bigl({\cal N}_T+\sqrt{5{\cal N}_T^2+4{\cal N}_T} \Bigr)
  \quad & {\rm if } \quad  & {\cal N}_T \le \frac{\sqrt{5}-2}{3}  \\
  \sqrt{2{\cal N}_T({\cal N}_T+1)}  - {\cal N}_T
  \quad & {\rm if } \quad  & {\cal N}_T \ge \frac{\sqrt{5}-2}{3}  \\
\end{array} \right.  .
\label{negfid}
\end{equation}
Upper bound for fidelity is realized for the family  
(\ref{sigmab}) or for any other state for which $C={\cal N}_T$.

Relative entropy of entanglement is bounded from above
by $E_F$. Numerical investigations suggest \cite{VADM01} 
that the lower bound is achieved for the family of (\ref{sigmac}),
which implies
\begin{equation}
E_F \ge E_R \ge \Bigl[ (C-2) \ln (1-C/2)+(1-C)\ln (1-C)\Bigr]
 \  .
\label{eforerel}
\end{equation}
Here $C^2=1-(2\mu_1-1)^2$ 
and $\mu_1=S^{-1}(E_F)$
denotes the larger of two preimages
of the entropy function (\ref{conc3}).
Similar bounds between relative entropy of entanglement,
and concurrence or negativity were studied in \cite{MG04}.

Making use of the analytical formulae 
for entanglement measures we may try to explore
the interior of the $15$-dim set of mixed states.
In general, the less quantum state is pure,
the less it is entangled: 
If $R=1/[{\rm Tr}\rho^2] \ge 3$ 
we enter the separable ball
and all entanglement measures vanish.
Also movements along an global orbit
$\rho \to U\rho\, U^{\dagger}$
generically changes entanglement.
For a given spectrum $\vec x$
the largest concurrence $C^*$,
which may be achieved on such an orbit
is given by Eq. (\ref{abssep4})
\cite{IH00,VAM01}.
Hence the problem of finding 
maximally entangled mixed states of two qubits
does not have a unique solution:
It depends on the measure of mixedness and
the measure of entanglement used. Both
quantities may be characterised e.g.
by a R{\'e}nyi entropy (or its function),
and for each choice of the pair of parameters $q_1,q_2$
one may find an extremal family of mixed states \cite{WNGKMV03}.

\vskip 0.45cm
\begin{figure} [tbp]
  \begin{center} \
 \includegraphics[width=12.8cm,angle=0]{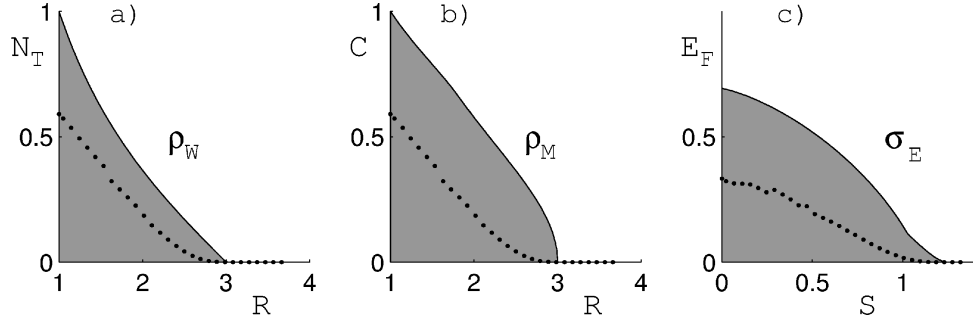}
\caption{Upper bounds for measures of entanglement as a function
of mixedness for two--qubits:
a) negativity  and  b) concurrence versus participation 
ratio $R=1/({\rm Tr}\rho^2)$;
c) entanglement of formation versus von Neumann entropy.
Gray shows entire accessible region while dots denote the
average taken with
 respect to the HS measure in ${\cal M}^{(4)}$. For pure states
it coincides with the average over FS measure.} 
\label{fig:ent09}
\end{center}
 \end{figure}

Figure \ref{fig:ent09} presents average entanglement plotted 
as a function of measures of mixedness
computed with respect to HS measure.
For a fixed purity ${\rm Tr} \rho^2$ 
the Werner states (\ref{Wernern})
 produce the maximal  negativity ${\cal N}_T$.
On the other hand, concurrence $C$ becomes
 maximal for the following states \cite{MJWK01}
\begin{equation}
\rho_M(y)  \equiv  
\left[
\begin{array}{c c c c}
 a & 0   & 0  & y/2 \\
 0   & 1-2a & 0  & 0 \\
 0   & 0   & 0 & 0 \\
 y/2  & 0   & 0 &  a
\end{array}
\right] ,
{\ \  \rm where \ \ }
\left\{
\begin{array}{ccc}
   a=1/3   \ \ & {\rm if } \ \   &  y \le 2/3  \\
   a=y/2   \ \  & {\rm if } \ \  &  y \ge  2/3  \\
\end{array} \right. 
\label{famM}
\end{equation}
Here $y \in [0,1]$ and  $C(\rho)=y$ while Tr$\rho^2=1/3+y^2/2$
in the former case 
and  Tr$\rho^2=1-2y(1-y)$ in the latter.
A family of states $\sigma_E$  
providing the upper bound of 
$E_F$ as a function of von Neumann entropy 
(see the line in Fig. \ref{fig:ent09}c) was found in  \cite{WNGKMV03}. 
Note that HS measure restricted to pure states
coincides with Fubini-Study measure.
Hence at $S=0$ the average
pure states entropy of entanglement
reads $\langle E_1\rangle _{\psi}=1/3$ while 
for $R=1$ we obtain $\langle C\rangle _{\psi}=$
$\langle {\cal N}_T \rangle_{\psi}= 3\pi/16\approx 0.59$.

\begin{figure} [htbp]
  \begin{center} \
 \includegraphics[width=9.9cm,angle=0]{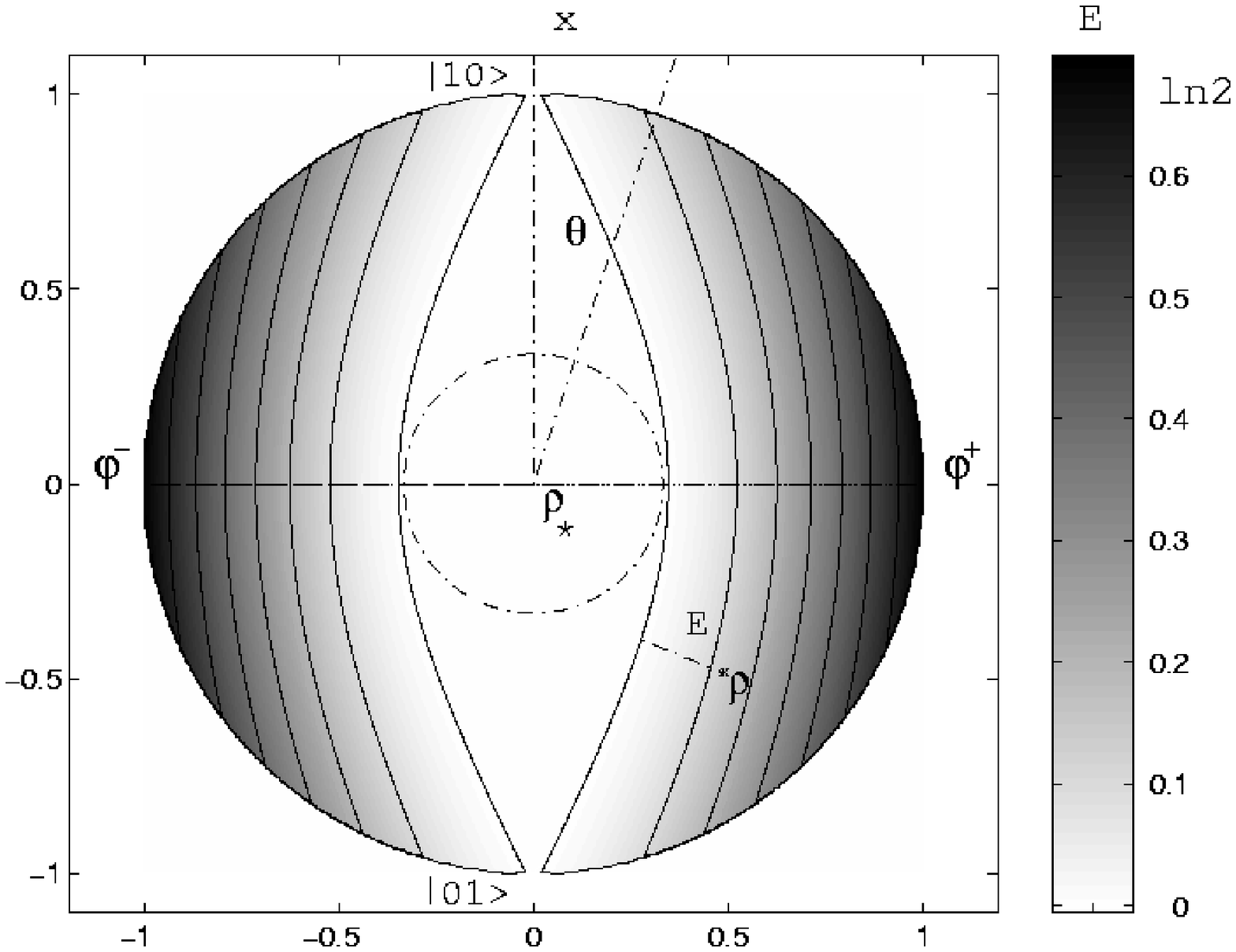}
\caption{Entanglement of formation
of generalised Werner states 
(\ref{efwewe})
in polar coordinates. White set represents
separable states.} 
\label{fig:ent10}
\end{center}
 \end{figure}

To close this section let us show  
in Fig. \ref{fig:ent10}
entanglement of formation for an
illustrative class of two-qubit states
\begin{equation}
\rho(x,\vartheta) \ \equiv \ 
x( |\psi_{\theta}\rangle \langle \psi_{\theta}|) +(1-x)\rho_*
{\rm \ \ with \ \ }
 |\psi_{\theta}\rangle = \frac{1}{\sqrt{2}}
\Bigl( \sin\frac{\vartheta}{2} |01\rangle 
+\cos \frac{\vartheta}{2} |10\rangle \Bigr)
 \  .
\label{efwewe}
\end{equation}
For $x=1$ the pure state is separable for $\vartheta=0,\pi$ and maximally
entangled $(*)$ for   $\vartheta=\pi/2,3\pi/2$. The dashed horizontal
line represents the Werner states.  The set ${\cal M}_S$ of separable states
 contains the maximal ball and touches the set of pure states in two
points. A distance $E$ of $\rho$ from the set ${\cal M}_S$
may be interpreted as a measure of entanglement.

We have come to the end of our tour across
the space of two qubit mixed states.
Since all the properties {\bf i)-x)}
break down for higher $N$,
 the geometry of quantum entanglement gets 
correspondingly more complex.
Already for the system of two qutrits 
the bound entangled states appear,
while the multi-partite problems
contain non-equivalent forms of
quantum entanglement.

\section{Concluding remarks}

\vspace{5mm}

The aim of this paper is to present  
literally to present
an introduction to the subject of
quantum entanglement. 
Although we have left untouched several important
 aspects of quantum entanglement, including multipartite systems, 
infinite dimensional systems, and continuous variables,
we hope that a reader may gain 
a fair overview of basic
properties of quantum entanglement.

What is such a knowledge good for?
We believe it will contribute to a 
better understanding of quantum mechanics.
We hope also that it will provide a solid foundation
for a new, emerging field of science--the
theory of quantum information processing.
Quantum entanglement plays a decisive role
in all branches of the  field 
including quantum cryptography,
quantum error correction, and quantum computing.

\smallskip

We would like to thank
Johan Br\"annlund, \AA sa Ericsson, 
Marek Ku\'s, Florian Mintert and Magdalena Sino{\l}{\c e}cka,
with whom we have worked on related projects.
We have also enjoyed constructive interactions with
Matthias Christandl, Jens Eisert, 
Micha{\l}, Pawe{\l} and Ryszard Horodeccy, Vivien Kendon,
Artur {\L}ozi{\'n}ski, Christian Schaffner, Paul Slater, 
and William Wootters.

It is our  pleasure to thank 
 for the hospitality at 
the Perimeter Institute for Theoretical Physics,
where this work  was done.
Financial support by Komitet Bada{\'n} Naukowych
under the grant PBZ-Min-008/P03/03,
VW grant 'Entanglement measures and the influence of noise'
and Swedish grant from VR is gratefully acknowledged.

\appendix

\section {Quantum entanglement: Further hints on the literature}
\label{cha:liter}

The number of papers devoted to quantum entanglement
keeps growing fast in the last decade.
For instance,  127 papers 
with the word {\sl entanglement} or {\sl entangled}
in the title were posted into the archives arXiv
during the first quarter of 2006 only.
We are not in position to present 
a review of all the papers which appeared after the book
was finished in March 2005. Instead we
mention in this appendix some exemplary references,
which make our story more complete.

The separability problem for $N \times N$
bi--partite systems remains open for$N\ge 3$.
Some spectral conditions implying separability
in such systems were provided in \cite{Ra06}.  
Efficient numerical algorithms for detection
of separability and entanglement were 
given in \cite{ITCE04,HBr04,BHH05}.

Valuable recent reviews on entanglement measures 
are provided by Plenio and Virmani \cite{PV05}
and Mintert et al. \cite{MCKB05}.
The monotonicity conditions for entanglement measures
have been simplified by  M. Horodecki \cite{Ho05},
while continuity bounds on the quantum relative entropy
were established in \cite{AE05}.
  
The logarithmic negativity 
was shown to be an entanglement monotone that is not convex
\cite{Pl05}.
Entanglement cost was proved by Yang et al. 
to be strictly larger than zero for any entangled state \cite{YHHSR05}.

Several versions of generalized concurrence for mixed quantum states
were investigated  in \cite{MB05,CAF05a,Go05,HBj05}.
Entanglement measures for rank-$2$ mixed states
where computed by Osborne \cite{Os05},
while lower bounds for entanglement of formation
of mixed states was found by Chen et al. \cite{CAF05b}.

Comparative analysis of various entanglement measures 
and the degree of mixing for two qubit systems
was performed in \cite{MG04}.
Entanglement witnesses for qubits and qutrits
were investigated in \cite{BDH05,JRY05}
while an application of witness operators 
to quantify the entanglement   
was proposed in \cite{Br05}.
Relations between entanglement witnesses and Bell inequalities
were discussed in \cite{HGBL05}.

Properties of the boundary of the set  of
mixed quantum states and its subset
containing separable states were investigated in \cite{GKM05}.
Later on the set of
two qubit separable states was proved
to have a constant width \cite{SBZ06}.
Geometric features of the set of entangled states
were analyzed in \cite{Le04b,LMO06}.

\medskip
\section{Contents of the book "Geometry of Quantum States. 
         An Introduction to Quantum Entanglement"
         by {\sl I. Bengtsson and K. {\.Z}yczkowski}}

{\small

 {\bf 1 Convexity, colours and statistics }
 
\quad {1.1} Convex sets                                

\quad {1.2} High dimensional geometry

\quad  {1.3} Colour theory

\quad  {1.4} What is ``distance''? 

\quad  {1.5} Probability and statistics

{\bf 2  Geometry of probability distributions } 

\quad {2.1} Majorization and partial order

\quad {2.2} Shannon entropy

\quad {2.3} Relative entropy

 \quad {2.4} Continuous distributions and measures

\quad {2.5} Statistical geometry and the Fisher--Rao metric

 \quad {2.6} Classical ensembles

\quad {2.7} Generalized entropies

 {\bf 3 Much ado about spheres}
 
\quad {3.1} Spheres

\quad  {3.2} Parallel transport and statistical geometry

 \quad {3.3} Complex, Hermitian, and K\"ahler manifolds

\quad  {3.4} Symplectic manifolds

\quad {3.5} The Hopf fibration of the 3-sphere

\quad  {3.6} Fibre bundles and their connections

\quad  {3.7} The 3-sphere as a group

\quad  {3.8} Cosets and all that

 {\bf 4  Complex projective spaces}

\quad  {4.1} From art to mathematics

\quad  {4.2} Complex projective geometry

\quad  {4.3} Complex curves, quadrics and the Segre embedding

\quad {4.4} Stars, spinors, and complex curves

\quad {4.5} The Fubini-Study metric

\quad {4.6} ${\mathbb C}{\bf P}^n$ illustrated

\quad {4.7} Symplectic geometry and the Fubini--Study measure

\quad {4.8} Fibre bundle aspects

\quad {4.9}Grassmannians and flag manifolds

 {\bf 5 Outline of quantum mechanics}

 \quad{5.1} Quantum mechanics

 \quad{5.2} Qubits and Bloch spheres

 \quad{5.3} The statistical and the Fubini-Study distances

 \quad{5.4} A real look at quantum dynamics

 \quad{5.5} Time reversals

 \quad{5.6} Classical \& quantum states: a unified approach

 {\bf 6 Coherent states and group actions }

 \quad{6.1} Canonical coherent states

 \quad{6.2} Quasi-probability distributions on the plane

 \quad{6.3} Bloch coherent states

 \quad{6.4} From complex curves to $SU(K)$ coherent states

 \quad{6.5} $SU(3)$ coherent states

 {\bf 7 The stellar representation}

 \quad{7.1} The stellar representation in quantum mechanics

 \quad{7.2} Orbits and coherent states

 \quad{7.3} The Husimi function

 \quad{7.4} Wehrl entropy and the Lieb conjecture

 \quad{7.5} Generalised Wehrl entropies

 \quad{7.6} Random pure states

 \quad{7.7} From the transport problem to the Monge distance

 {\bf 8 The space of density matrices}

 \quad{8.1} Hilbert--Schmidt space and positive operators

 \quad{8.2} The set of mixed states

 \quad{8.3} Unitary transformations

 \quad{8.4} The space of density matrices as a convex set

\quad{8.5} Stratification

 \quad{8.6} An algebraic afterthought

 \quad{8.7} Summary

 {\bf 9 Purification of mixed quantum states}

 \quad{9.1} Tensor products and state reduction

\quad{9.2} The Schmidt decomposition

\quad {9.3} State purification \& the Hilbert-Schmidt bundle

\quad {9.4} A first look at the Bures metric

 \quad{9.5} Bures geometry for $N=2$

 \quad{9.6} Further properties of the Bures metric

 {\bf 10 Quantum operations} 

 \quad{10.1} Measurements and POVMs

 \quad{10.2} Algebraic detour: matrix reshaping and reshuffling

 \quad{10.3} Positive and completely positive maps

\quad {10.4} Environmental representations

\quad {10.5} Some spectral properties

\quad{10.6} Unital \& bistochastic maps

\quad {10.7} One qubit maps

 {\bf 11 Duality: maps versus states}

 \quad{11.1} Positive \& decomposable maps

\quad {11.2} Dual cones and super-positive maps

 \quad{11.3} Jamio{\l }kowski isomorphism

 \quad{11.4} Quantum maps and quantum states

 {\bf 12  Density matrices and entropies} 

 \quad{12.1} Ordering operators

 \quad{12.2} Von Neumann entropy

\quad {12.3} Quantum relative entropy

\quad {12.4} Other entropies

 \quad{12.5} Majorization of density matrices

\quad {12.6} Entropy dynamics

 {\bf 13 Distinguishability measures}

\quad{13.1} Classical distinguishability measures

\quad {13.2} Quantum distinguishability measures

\quad {13.3} Fidelity and statistical distance

 {\bf 14 Monotone metrics and measures}

\quad {14.1} Monotone metrics

\quad {14.2} Product measures and flag manifolds

 \quad{14.3} Hilbert-Schmidt measure

 \quad{14.4} Bures measure

 \quad{14.5} Induced measures

 \quad{14.6} Random density matrices

\quad {14.7} Random operations

 {\bf 15 Quantum entanglement}

 \quad{15.1} Introducing entanglement

 \quad{15.2} Two qubit pure states: entanglement illustrated

 \quad{15.3} Pure states of a bipartite system

 \quad{15.4} Mixed states and separability

\quad {15.5} Geometry of the set of separable states

 \quad{15.6} Entanglement measures

 \quad{15.7} Two qubit mixed states 

{\bf Epilogue }

 \quad  Appendix 1 Basic notions of differential geometry 

\quad Appendix 2 Basic notions of group theory 

\quad Appendix 3  Geometry do it yourself 

\quad Appendix 4  Hints and answers to the exercises
} 

\medskip

\section {\ Geometry---do it yourself}
\label{cha:extra}

\noindent In this appendix we provide some additional
 exercises of a more practical nature.
 Working them out might improve our skills and intuition
 required to cope with multi-dimensional geometric objects 
 which arise while investigating quantum entanglement.

 \medskip

\noindent {\bf Exercise 1 -- Real projective space}.
\index{space!real projective} 
Cut out a disk of radius $r$. Prepare a narrow strip of 
length $\pi r$ and glue it into a M{\"o}bius strip.
The total length of the boundary of a strip
is equal to the circumference of the disk
so you may try to glue them together.
(If you happen to work
in $3$ dimensions this simple task gets difficult...)
If you are done,  you can contemplate a nice model of 
a real projective space, ${\mathbbm R}{\bf P}^2$.
\vskip -0.3cm 
\begin{figure}[htbp]
        \centerline{ \hbox{
                \epsfig{figure=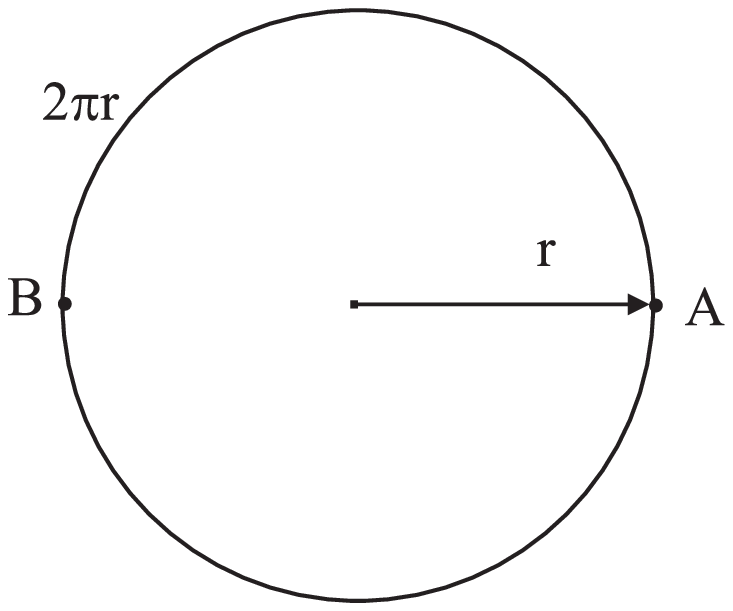, width=3.72cm} 
   }}
\end{figure}
\vskip -0.8cm 
\begin{figure}[htbp]
        \centerline{ \hbox{
                \epsfig{figure=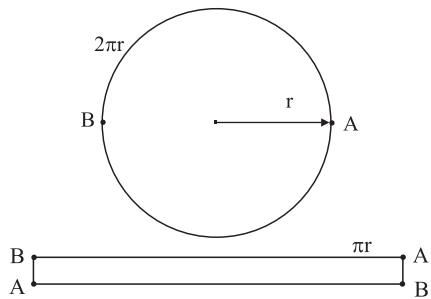, width=5.6cm} 
   }}
        \caption{A narrow M{\"o}bius strip glued with a circle produces 
          ${\mathbb R}{\bf P}^2$.}
        \label{fig:extra1}
\end{figure}

\begin{figure}[htbp]
        \centerline{ \hbox{
             \epsfig{figure=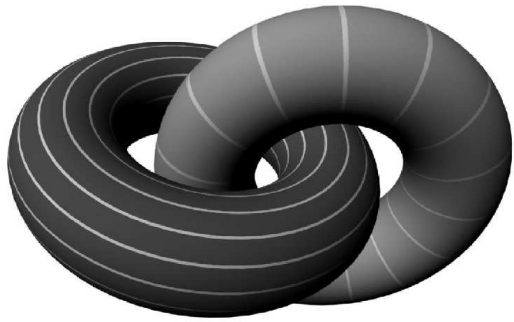, width=70mm}}}
        \caption{Heegard decomposition of a three--sphere.} 
        \label{fig:extra2}
\end{figure}

 \smallskip 
\noindent {\bf Exercise 2 -- Hypersphere}
 ${\bf S}^3$ may be obtained by identifying 
points on the surfaces of two identical $3$--balls
as discussed in section 3.3
To experience further features of the 
hypersphere get some playdough and prepare  
two cylinders of different colours with their length more than three times 
larger than their  diameter.
Form two linked tori as shown in Fig.  \ref{fig:extra2}.

Start gluing them together along their boundaries. 
After this procedure (not easy in 3-D)  is completed 
you will be in position to astonish your
colleagues by presenting them a genuine
\index{sphere!Heegard decomposition} 
{\it Heegard decomposition} of a hypersphere. 
%

\begin{figure}[htp]
        \centerline{ \hbox{
          \epsfig{figure=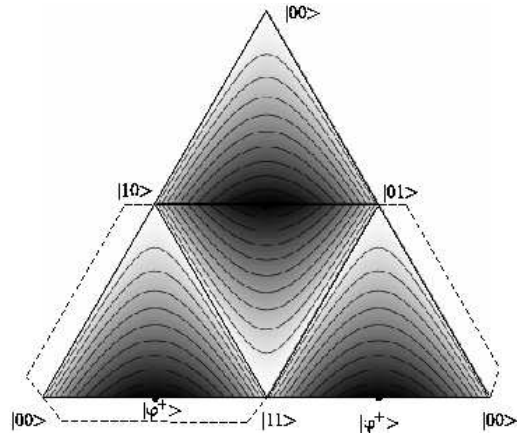,width=7.2cm}}}
        \caption{Net of the tetrahedron representing
   entanglement for pure states of two qubits:
   maximally entangled states plotted in black.}
  \label{fig:extra4}
\end{figure}

 \smallskip 
\medskip
\noindent {\bf Exercise 3 -- Entangled pure states}.
Magnify Fig. \ref{fig:extra4} and cut out the net of the cover tetrahedron.
It represents the entanglement of formation
of the pure states of two qubits for  
a cross-section of ${\mathbb C}{\bf P}^{3}$. 

Glue it together to get the entanglement tetrahedron with four 
product states in four corners. 
Enjoy the symmetry of the object and 
study the contours of the states
of equal entanglement.

\begin{figure}[htbp]
        \centerline{ \hbox{
                \epsfig{figure=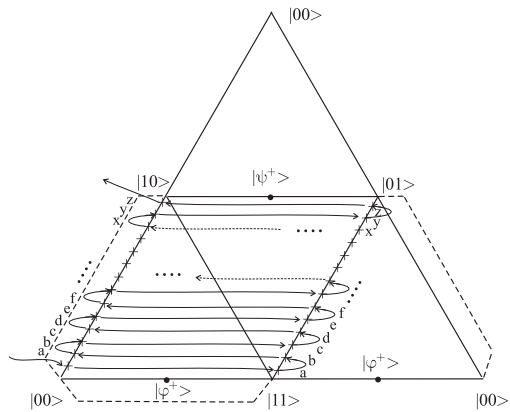,width=6.8cm}}}
        \caption{Sew with a colour thread inside a transparent tetrahedron
    to get the ruled surface consisting of separable pure states
of two qubits.}
 \label{fig:extra5}
        \end{figure}

 \smallskip
\noindent {\bf Exercise 4 -- Separable pure states}.
Prepare a net of a regular tetrahedron from 
transparency according to the blueprint shown in Fig. \ref{fig:extra5}.
Make holes with a needle along two opposite edges as shown in the picture.
Thread a needle with a (red) thread and start sewing it through your model.
Only after this job is done glue the tetrahedron together. (Our practice
shows that sewing after the tetrahedron is glued together is much more
difficult...)
If you pull out the loose tread
and get the object sketched in Fig. \ref{fig:Oktfig6},
you can contemplate, how a fragment of the subspace of separable states
forms a ruled surface embedded inside the tetrahedron. 

\medskip


\end{document}